\documentclass{aa}
\usepackage{txfonts}
\usepackage{graphicx}
\usepackage{natbib}
\bibpunct{(}{)}{;}{a}{}{,}

\begin{document}
%\onecolumn

\title{Evolution of warped and twisted accretion discs in close binary
systems}
\author{Moritz M. Fragner \& Richard P. Nelson.}
\institute{Astronomy Unit, Queen Mary, University of London, Mile
End Road, London E1 4NS.\\
\email{M.Fragner@qmul.ac.uk, R.P.Nelson@qmul.ac.uk}}
\date{Received/Accepted}

\abstract
% Context:
{There are numerous examples of accretion discs in binary
systems where the disc midplane is believed to be inclined
relative to the binary orbit plane.}
% Aims:
{We aim to examine the detailed disc structure that
arises in a misaligned binary system as a function of the disc
aspect ratio $h$, viscosity parameter $\alpha$, disc outer radius $R$,
and binary inclination angle $\gamma_F$. We also aim to examine the
conditions that lead to an inclined disc being disrupted by strong differential
precession.}
% Methods:
{We use a grid-based hydrodynamic code to perform 3D simulations. 
This code has a relatively low numerical viscosity
compared with the SPH schemes that have been used previously to
study inclined discs. This allows the influence of viscosity 
on the disc evolution to be tightly controlled. }
% Results:
{We find that for thick discs ($h=0.05$) with low $\alpha$, 
efficient warp communication in the discs allows them
to precess as rigid bodies with very little warping
or twisting. Such discs are observed to align with the
binary orbit plane on the viscous evolution time.
Thinner discs with higher viscosity, in which warp communication is 
less efficient, develop significant twists before achieving a 
state of rigid-body precession. Under the most extreme conditions
we consider ($h=0.01$, $\alpha=5 \times 10^{-3}$ and $\alpha=0.1$),
we find that discs can become broken or disrupted by strong differential
precession. Discs that become highly
twisted are observed to align with the binary orbit plane on
timescales much shorter than the viscous timescale, possibly
on the precession time.}
%Conclusions:
{We find agreement with previous studies that show that thick
discs with low viscosity experience mild warping and precess rigidly.
We also find that as $h$ is
decreased substantially, discs may be disrupted by
strong differential precession, but for disc thicknesses that are
significantly less ($h=0.01$) than those found in previous studies
($h=0.03$).}

\titlerunning{Warped Accretion Discs in Binary Star Systems}
\authorrunning{M.M. Fragner \& R.P. Nelson}

\maketitle
\section{Introduction}\label{intro}
Accretion discs are found in a number of astrophysical environments. 
These include protostellar discs around T Tauri stars,
and around compact objects in cataclysmic variable and X-ray binary
systems. 
Most T Tauri stars associated with low-mass star-forming regions 
are observed  to be members of binary systems, where the
peak in the distribution of separations occurs at
approximately 30 AU \citep{leinert}. Studies of the tidal
interaction between a circumstellar disc and a binary companion
indicate that tidal truncation occurs at a ratio of binary
separation to disc radius $D/R \simeq 3$ for an equal mass binary \citep{arty,larwood,kleynelson}.
Given that 30 AU is 
smaller than typical disc sizes observed in T Tauri systems \citep{edwards}, 
protostellar discs in many binary systems are often going to experience
strong tidal effects.

There are observational indications that the disc and binary orbital 
plane are not always coplanar. The binary system HK Tau has been
imaged, and shows compelling evidence for a disc whose
midplane is misaligned with the orbit of the binary companion
\citep{stapelfeldt}.
\cite{terquem1, terquem2} have suggested that the reprocessing of radiation 
from the central star by a tilted and warped accretion disc could 
account for the high spectral index of some T Tauri stars.
A number of precessing jets have been observed in star forming
regions, and the jet precession may indicate an
underlying precessing disc from which the jet is launched 
\citep{koenig,eisloffel}.
In the case of close X-ray binaries, the objects SS433 and Her X-1 
show evidence of precessing jets and/or accretion discs 
\citep{boynton,katz,margon,petterson}.

An early study of the equations describing the behaviour of a 
thin non-coplanar viscous disc was undertaken by \cite{petterson}. 
The resulting equations, however, lacked the property of 
conserving global angular momentum. This was addressed in 
subsequent work by \cite{pringle}, where they showed 
that warps in discs for which $h < \alpha < 1$ evolve diffusively on 
a timescale shorter than the usual viscous accretion by a factor 
$\alpha^2$, where $\alpha$ is the dimensionless 
viscosity coefficient \citep{shakura}, and $h$ is the disc
aspect ratio $H/r$.
The opposite regime $\alpha\le h$ has been investigated by 
\cite{paplin} using linear perturbation theory. They demonstrated 
that the governing equation for the disc tilt changes 
from being a diffusion equation to a wave equation 
in this physical regime, showing that warps propagate 
as bending waves with a speed corresponding to half the sound speed, $c_s/2$.
An analytic study of warped discs was undertaken
by \cite{ogilvie} in which the mildly non linear evolution of disc
warps was examined.

The tidal perturbation of an inviscid disc by a companion 
on an inclined circular orbit has been studied by 
\cite{papterquem} using linear theory.
They investigate the disc response to tidal perturbations that 
have odd symmetry with respect to the disc midplane and 
consider both zero and non-zero perturbing frequencies.
They show that the zero-frequency (secular) mode leads 
to rigid precession of the disc if the sound crossing time through 
the disc is smaller than the differential precession time.
Furthermore, they estimate that the timescale for damping the 
inclination is of the same order as the accretion timescale of the disc.
A study of the twisting and warping of viscous discs in binary systems was 
undertaken by \citet{lubow} using linear theory, and it was shown that discs 
whose outer disc radii are smaller than the truncation radius maybe unstable to tilting.

Numerical simulations of tidally interacting discs which are not 
coplanar with the binary orbit have been performed by 
\cite{larwood} using SPH simulations. They found that the 
disc precesses approximately as a rigid body as long as the 
disc aspect ratio is large enough. For smaller aspect ratios their 
results suggest that the disc develops a modest warp, while for 
$h\le 0.03$ the disc becomes disrupted as a consequence 
of differential precession. Their results suggest that a disc can 
split-up into two distinct parts that behave independently.
These SPH simulations also showed that the inclination evolves 
on the viscous timescale, as expected from \cite{papterquem}. 

The goal of the present study is to examine in detail
the structure of misaligned accretion discs in close binary
systems as a function of the important physical parameters:
$h$, $\alpha$, the outer disc radius $R$, and the inclination
angle $\gamma_F$. The binary companion and the central star are assumed to be of equal mass.
We use a grid-based code (NIRVANA) to perform
this study, and this code has a relatively small numerical
viscosity compared with the SPH schemes used to perform
previous studies. This allows us to tightly control the
disc viscosity, and therefore examine in more detail
its influence on the disc evolution.
We consider a broad range of disc parameters in which
warps propagate either as bending waves $h > \alpha$ or
diffusively $h < \alpha$. As well as examining the 
quasi-steady disc structures which arise because of tidal interaction
with the inclined companion, we also examine the conditions
under which a disc becomes disrupted due to strong differential
precession. In particular we are interested in examining whether
the following criterion holds: a disc will achieve a state
of rigid-body precession if warp propagation across the disc
(either through bending waves or diffusion) occurs on a timescale
shorter than the differential precession time.

The plan of the paper is as follows. In Sect.~2 we present the 
basic equations of the problem. In Sect.~3 we describe the numerical 
methods used in the simulations, and in Sect. 4 we discuss 
the linear theory of bending waves and calibrate the simulation code
against calculations based on linear theory.
In Sect.~5 we present our results for the tilted, tidally 
interacting discs, and in Sect. 6 we discuss our results and
present our conclusions.

\section{Basic equations}
\label{Basic-equations}

We consider the evolution of a thin, gaseous, viscous disc in orbit
around a central star, where the disc is perturbed by a
companion star orbiting in a plane which is not coincident
with the disc midplane. For a range of physical parameters, it is expected
that the disc will precess rigidly around the angular
momentum vector of the binary system.
The equations of continuity and motion for a viscous fluid
in a precessing reference frame may be written as:

\begin{equation}
\frac{D\rho}{Dt}+\rho {\bf \nabla} . {\bf v}=0
\label{continuity}
\end{equation}

\begin{equation}
\frac{D{\bf v}}{Dt}=-\frac{1}{\rho}{\bf\nabla} P-{\bf\nabla}\Psi+{\bf S}_{visc}-2{\bf\Omega}\times{\bf v}-{\bf\Omega}\times ({\bf\Omega}\times{\bf r})
\label{momentum}
\end{equation}
where
\begin{eqnarray}
\frac{D}{Dt}=\frac{\partial}{\partial t}+{\bf v}\cdot{\bf\nabla}\nonumber
\end{eqnarray}
is the convective derivative, $\rho$ is the density,
${\bf v}$ the velocity and P the pressure. The precession frequency is
given by $| {\bf \Omega}|$, and the disc angular momentum vector is
assumed to precess around a vector pointing in the direction of ${\bf \Omega}$.
The gravitational potential is $\Psi$.
We adopt a locally isothermal equation of state:
\begin{eqnarray}
P=c_s(r)^2\rho
\end{eqnarray}
where the isothermal sound speed is defined by
$c_s(r)=h\cdot v_{Kep}$. Here $h$ is the
aspect ratio of the disc, $H/r$, and $v_{Kep}$ is the 
local keplerian velocity. ${\bf S}_{visc}$ is the viscous
force per unit mass:
\begin{eqnarray}
{\bf S}_{visc}=\frac{1}{\rho}{\bf\nabla}\underline{\underline{T}}
\end{eqnarray}
where $\underline{\underline{T}}$ is the viscosity stress tensor:
\begin{eqnarray}
T_{ij}=\rho \nu \left(\frac{\partial v_i}{\partial x_j} +
\frac{\partial v_j}{\partial x_i} - 
\frac{2}{3}\delta_{ij}{\bf\nabla}.{\bf v}\right)
\end{eqnarray}
We adopt the standard `alpha' model (Shakura \& Sunyaev 1973) to
specify the kinematic viscosity $\nu = \alpha c_s H$.
We solve the above equations numerically, using a system of
spherical coordinates ${\bf r}=$ ($r$, $\theta$, $\phi$).

\subsection{Orbital configuration}
We work in a reference frame in which the origin of the coordinate system
is fixed on the central star, and the secondary star moves
on a circular orbit about the origin with position vector ${\bf D}$.
The gravitational potential, $\Psi$, at any position vector
${\bf r}$ is therefore given by:
\begin{eqnarray}
\Psi=-\frac{GM_P}{r}-\frac{GM_S}{|{\bf r}-{\bf D}|}+
\frac{GM_S{\bf r}\cdot{\bf D}}{D^3}, 
\end{eqnarray}
where $G$ is the gravitational constant, and $M_P$ and $M_S$ are the primary
and secondary masses, respectively. The first term is due to the
primary star, and the remaining two terms are due to the companion star.
The last indirect term accounts for the acceleration of the origin of
the coordinate system, which coincides with the location of the primary star.
Note that there are no contributions to the potential due to the disc
since we do not include disc self-gravity in our calculations,
and the stars do not experience a gravitational force due to the disc.
Consequently the binary orbital parameters remain unchanged.

Hence the secondary star feels the acceleration due to the central star
and the indirect component of the potential.
%, such that the equation
%for ${\bf D}$ may be written
%\begin{eqnarray}
%\frac{d^2{\bf D}}{dt^2}=-\frac{GM_P}{D^3}{\bf D}-\frac{GM_S}{D^3}{\bf D}=-\omega^2{\bf D}.
%\end{eqnarray}
In a non-precessing reference frame, centred on the central star, 
whose $x-y$ plane coincides with the orbital plane of the binary,
the position vector of the companion in a circular orbit may be written as:
\begin{eqnarray}
{\bf D}=D\cdot(\cos{(\omega t)}{\bf e}_1+\sin{(\omega t)}{\bf e}_2),
\label{sec_solution}
\end{eqnarray}
where $D$ is the constant binary separation and $\omega=\sqrt{G(M_P+M_S)/D^3}$ is the binary
orbital angular velocity.

\section{Numerical methods}
The system of equations described in Sect.~\ref{Basic-equations}
is integrated using the grid-based hydrodynamics
code NIRVANA \citep{ziegler}, adapted to solve the equations 
in a precessing reference frame. This code uses operator-splitting,
and the advection routine uses a second-order 
accurate monotonic transport algorithm \citep{leer}.

\subsection{Boundary conditions}
Periodic boundary conditions were applied in the azimuthal
direction. At all other boundaries outflow conditions were adopted
using zero-gradient extrapolation. Velocities at the radial boundaries
were set to ensure that zero viscous stress occurs there.

\subsection{Units}
We adopt a system of units such that the unit of mass is
equal to that of the central star, $M_P$, the unit
of radius is equal to the radius of the inner boundary
of our computational domain, and we set the gravitational constant $G=1$.
The unit of time then becomes $1/\Omega_K(1)$, where $\Omega_K$ is
the keplerian angular velocity evaluated at $r=1$.
When discussing simulation results, however, we will express time
in units of the orbital period at $r=10$, the nominal outer disc
radius in the simulations. Thus we describe a time interval of
$2\pi\Omega_K(10)^{-1}$ as `one orbit'.
Inclination and precession angles are described in units of degrees.

\subsection{Reference frames and initial conditions}
The numerical domain extends radially from $r=[1,R]$, 
and azimuthally from $\phi=[0,360^\circ]$.
The outer radius, $R$, may differ from simulation to simulation.
The meridional domain covers the range
$\theta=[90^\circ-\Delta\theta,90^\circ+\Delta\theta]$, where again
$\Delta\theta$ may vary between the simulations.

We perform simulations in two different reference frames.
For models in which the disc is expected to develop rigid precession,
we solve the equations in a frame which precesses around the
angular momentum vector of the binary.  Given a sensible choice of
the precession rate, this approach ensures that the disc
midplane always stays close to the equatorial plane of the
computational domain where $\theta=90^\circ$, and allows
simulations to be performed with large inclinations of the
binary orbit. If evolved in a non-precessing frame, the disc
would precess around the binary angular momentum vector, causing it
to eventually intersect with the upper and lower meridional boundaries of
the computational domain if the binary inclination angle is large.
Adopting a precessing reference frame allows simulations to be
conducted with relatively small meridional domains, even for
large binary inclinations, thus substantially reducing the 
computational expense involved.

For models in which substantial differential precession of the disc
is expected to arise, adopting a frame with a single
precession frequency does not solve the problem described above,
since some parts of the disc inevitably intersect the meridional boundaries.
As these models are of significant interest from the 
astrophysical point of view (they will tend to be thin discs
with large viscosity, similar to those which occur in X-ray binaries
for example),
we have computed them in a non precessing frame, but have been forced
to use large meridional domains and smaller binary inclination angles.

We denote coordinates in the precessing frame as ${\bf\hat{r}}$,
while coordinates in the non-precessing binary frame are denoted ${\bf r}$.
The coordinates in the code frame are denoted with ${\bf r}_C$,
where the code frame can be one of either the precessing frame or the
binary frame.
The transformation of any vector ${\bf e}$ from the binary into
the precessing frame is given by:
\begin{eqnarray}
{\bf \hat{e}}=R_Z(\Omega_F t)R_X(\gamma_F){\bf e}
\label{transform}
\end{eqnarray}
with rotation matrices:
\begin{eqnarray}
R_X(\gamma_F)&=&\left(\begin{array}{ccc}1&0&0\\0&\cos(\gamma_F)&\sin(\gamma_F)
\\0&-\sin(\gamma_F)&\cos(\gamma_F)\end{array}\right)\nonumber\\
R_Z(\Omega_F t)&=&\left(\begin{array}{ccc}\cos(\Omega_F t)&-\sin(\Omega_F t)&
0\\ \sin(\Omega_F t)& \cos(\Omega_F t)&0\\0&0&1\end{array}\right).
\end{eqnarray}
Thus to transform a vector from the binary into the precessing frame,
we first rotate around the $x$-axis by an inclination angle
$\gamma_F$ and then rotate the resulting vector around the
$z$-axis by an precession angle $\Omega_F t$. Here $\Omega_F$ is
the precession rate of the precessing frame and $t$ is the time elapsed.
This transformation
is particularly useful when setting up the initial conditions for the
simulations which are performed in the non-precessing binary frame,
since we assume that the binary orbit lies in a plane which
is coincident with the equatorial plane of the computational grid,
with the disc being inclined by an angle $\gamma_F$.

\subsubsection{Model set-up in precessing frame}
In the precessing frame the code midplane (defined to reside where 
$\theta_C=90^{\circ}$) coincides with the disc midplane
initially, so that  (${\bf r}_C \equiv {\bf\hat{r}}$).
The initial density profile is chosen to satisfy hydrostatic
equilibrium in the $\hat{\theta}$-direction, which yields the result:
\begin{eqnarray}
\rho(r,\hat{\theta})=r^{-\zeta}\left[
\sin{(\hat{\theta})}\right]^{\left(\frac{1}{h^2}-(\zeta+1)\right)},
\label{density}
\end{eqnarray}
where $\zeta=\frac{3}{2}$ is the radial power law
exponent of the density profile.
Note that this expression has the limit:
\begin{eqnarray}
\lim_{\hat{\theta}\rightarrow 0} \;\;  \rho(r,\hat{\theta})=r^{-\zeta}
exp\left(-\frac{\left(\hat{\theta}-\frac{\pi}{2}\right)^2}{2h^2}\right),
\end{eqnarray}
which is the familiar form of the density profile for thin discs.
The velocity in the $\hat{\phi}$ direction is chosen to balance the
forces in the radial direction:
\begin{eqnarray}
\hat{v_\phi^2}=\frac{1}{r}(1-(\zeta+1)h^2).
\label{phivelocity}
\end{eqnarray}
Consequently, the azimuthal velocity profile is slightly subkeplerian
because of the radial pressure gradient.
The velocities in the $r$ and $\hat{\theta}$ direction are
chosen to be zero initially.\\
We set ${\bf \Omega}$ in Eq.~(\ref{momentum}) to point in the same direction
as the binary angular momentum vector:
\begin{eqnarray}
{\bf\Omega}=\Omega_F {\bf e}_3=\Omega_F(-\sin{(\gamma_F)}
{\bf\hat{e}}_2+\cos{(\gamma_F)}{\bf\hat{e}}_3)
\end{eqnarray}
where the last equality is derived using the transformation given
by Eq.~(\ref{transform}).
The precession rate may be estimated using the expression:
\begin{eqnarray}
\Omega_F=-\left(\frac{3GM_S}{4D^3}\right)
\frac{\int_0^R \Sigma r^3dr}{\int_0^R \Sigma r^3\Omega dr}
\cos{(\gamma_F)},
\end{eqnarray}
where $\Sigma=\int_{-\infty}^\infty\rho \, r \, d\theta$ 
is the disc surface density.
Hence the precession frequency is determined by the ratio
of the integrated external torque exerted by the secondary to the
integrated angular momentum content of the disc \citep{papterquem}.
For the velocity and density profiles given by Eqs.~(\ref{phivelocity})
and (\ref{density}), respectively, this evaluates to:
\begin{eqnarray}
\frac{\Omega_F}{\Omega_d(R)}=-\frac{3}{7}\frac{R^3}{D^3}
\frac{M_S}{M_P} \cos{(\gamma_F)},
\label{omega}
\end{eqnarray}
where $\Omega_d(R)$ is the orbital angular velocity of 
the disc at radius $R$.
The binary orbit given by Eq.~(\ref{sec_solution}),
expressed in precessing frame coordinates, is given by:
\begin{eqnarray}
{\bf D}=D\left(\begin{array}{c}\cos{([\omega-\Omega_F]t)}\\
\sin{([\omega - \Omega_F]t)} \cos(\gamma_F)\\
\sin{([\omega-\Omega_F]t))} \sin{(\gamma_F)}
\end{array} \right).
\end{eqnarray}
Thus an observer moving with the disc sees an increased binary frequency
$\omega -\Omega_F$ due to the retrograde precession of the disc
(i.e. $\Omega_F$ is negative).

\subsubsection{Model set-up in binary frame}
When working in the non-precessing binary frame we set the precession
frequency ${\bf\Omega}= 0 $ in Eq.~(\ref{momentum}).
Since the equatorial plane of the computational domain
coincides with the binary orbit plane (${\bf r}_C \equiv {\bf r}$),
the disc is inclined with respect to the equatorial plane of
the computational grid by an inclination angle $\gamma_F$. We can use the
transformation defined by Eq.~(\ref{transform}) at $t=0$ to
relate the polar angles of the precessing frame to the
polar angles of the binary frame. One particularly useful relation is given by:
\begin{eqnarray}
\sin^2{(\hat{\theta})} &=& \cos^2{(\phi)} \sin^2{(\theta)}
+ \cos^2{(\gamma_F)} \sin^2{(\phi)} \sin^2{(\theta)} \nonumber  \\ &+&
\sin^2{(\gamma_F)} \cos^2{(\theta)} \nonumber \\ &-&
2 \cos^2{(\gamma_F)} \sin{(\gamma_F)} \sin{(\theta)}
\cos{(\theta)} \sin{(\phi)}
\label{sintheta}
\end{eqnarray}
which can be used to determine the initial density profile given
by Eq.~(\ref{density}).
The initial velocity is given by:
\begin{eqnarray}
{\bf\hat{v}}_{\phi} &=& \hat{v}_\phi (-\sin{(\hat{\phi})}{\bf \hat{e}}_1
+ \cos{(\hat{\phi})}{\bf \hat{e}}_2) \nonumber \\ &=&
\hat{v}_\phi(-\sin{(\hat{\phi})}{\bf e}_1+\cos{(\hat{\phi})}
\cos{(\gamma_F)}{\bf e}_2 - \cos{(\hat{\phi})} \sin{(\gamma_F)}
{\bf e}_3)\nonumber
\end{eqnarray}
Using the relationship between the polar angles one can find
an expression for ${\bf\hat{v}}_\phi$ that is written entirely
in terms of binary frame angles. The different components
in the radial, azimuthal and meridional direction are then given by:
\begin{eqnarray}
&v_r&=0\nonumber\\
&v_\phi&=\hat{v}_\phi \frac{\cos{(\gamma_F)} \sin{(\theta)}-
\sin{(\phi)} \sin{(\gamma_F)} \cos{(\theta)}}{\sin{(\hat{\theta})}}
\nonumber\\
&v_\theta& =\hat{v}_\phi \frac{\sin{(\gamma_F)}
\cos{(\phi)}}{\sin{(\hat{\theta})}}
\label{tiltvelocity}
\end{eqnarray}
where $\hat{v}_{\phi}$ is given by Eq.~(\ref{phivelocity}) and
$\sin{(\hat{\theta})}$ by Eq.~(\ref{sintheta}).
It can be shown that this set of profiles satisfies the Euler equations
given by Eqs.~(\ref{continuity}) and (\ref{momentum}) with
${\bf S}_{visc}=0$, $M_S=0$, ${\bf\Omega}=0$ and $\gamma_F={\rm constant}$,
since in a spherically symmetric potential there is no preferred
direction for the disc rotation axis.

\subsection{Definition of precession and inclination angles}
In order to follow the evolution of the disc structure we calculate
the total angular momentum vector, ${\bf L}(r)$, for disc material 
contained in each spherical shell in the numerical domain:
\begin{eqnarray}
{\bf L}(r) &=& \int_0^{2\pi} d \phi \int_{\frac{\pi}{2}-\Delta\theta}^{\frac{\pi}{2}+
\Delta \theta} {\bf l} \cdot r^2 \, \sin{(\theta)} \, d\theta \, \delta r
\label{angmom}
\end{eqnarray}
where $\delta r$ is the radial grid spacing and ${\bf l}$ is the
angular momentum density:
\begin{eqnarray}
{\bf l}=\rho {\bf r}\times{\bf v}=\rho (-r v_\phi {\bf\theta}_C
+ r v_\theta {\bf\phi}_C).
\label{angmomdensity}
\end{eqnarray}
${\bf \theta}_C$ and ${\bf\phi}_C$ are the code unit vectors in
the meridional and azimuthal directions, respectively.
Note that when working in the precessing frame a term
involving the precessional velocities should be added to
this expression. In practice we find that this is of negligible
magnitude, because of the slow rate of precession, and therefore 
the term has been omitted.
For disc material located in a given spherical shell
we locate the inclination angle,
$\delta$, (equal to $\gamma_F$ at time $t=0$) between the disc and
binary angular momentum vectors through:
\begin{eqnarray}
\cos{(\delta)}=\frac{{\bf J}_B . {\bf L}}{|{\bf J}_B||{\bf L}|}.
\label{inclination}
\end{eqnarray}
where ${\bf J}_B$ represents the binary angular momentum vector.
The precession angle, $\beta$, measured in the binary plane can
be defined through:
\begin{eqnarray}
\cos{(\beta)}=\frac{({\bf J}_B \times{\bf L}).{\bf u}}{|{\bf J}_B
\times {\bf L}|},
\label{precession}
\end{eqnarray}
where ${\bf u}$ is an arbitrary reference unit vector in the binary plane.
We choose ${\bf u}=-{\bf e}_1$, so that the initial precession angle
$\beta=0$ by this definition.
When working in the precessing frame $-{\bf e}_1$ has to be expressed
in precessing frame coordinates and becomes time dependent.

When discussing the results of the simulations we use the
the following terminology: a {\it warp} occurs if the angle
$\delta$ becomes a function of radius in the disc; a disc develops
a {\it twist} when $\beta$ becomes a function of radius; a disc
is said to be {\it broken} if either $\beta$ or $\delta$
change discontinuously at some radius such that the disc
separates into two independently precessing parts; 
a disc is said to be {\it disrupted} if $\beta$ varies smoothly
by more than 180 degrees across the disc radius because of
differential precession.

\subsection{Calculation of torque contributions}
Later in this paper we will be interested in measuring the different
contributions to the changes in the precession and inclination angles
for disc material located at different radii in the computational domain.
As a first step we calculate the rate of change of the angular momentum vector
${\bf L}(r)$ associated with disc material located at radius $r$,
which involves integrating over the individual spherical shells
in the computational domain.

Using the continuity and momentum Eqs.~(\ref{continuity}) and
(\ref{momentum}) to express the velocity and density changes in
Eq.~(\ref{angmomdensity}), and assuming that the density vanishes at
the meridional boundaries, we find the result:
\begin{eqnarray}
{\bf\dot{L}}(r)={\bf\dot{L}}^F(r)+{\bf\dot{L}}^V(r)+{\bf\dot{L}}^S(r)
\label{torques}
\end{eqnarray}
where a dot denotes a time derivative. We have that
\begin{eqnarray}
{\bf\dot{L}}^{F}(r)=-\int_0^{2\pi}d\phi\int_{\frac{\pi}{2}-\Delta\theta}^{\frac{\pi}{2}+\Delta\theta}d\theta \sin{(\theta)} \left[r^2\cdot{\bf l}\cdot v_r\right]^{r+\frac{\delta r}{2}}_{r-\frac{\delta r}{2}}
\label{flux}
\end{eqnarray}
is the change due to the divergence of a radial angular momentum flux.
The change due to viscous interaction with neighboring shells is given by:
\begin{eqnarray}
{\bf\dot{L}}^{V}(r)=\int_0^{2\pi}d\phi\int_{\frac{\pi}{2}-\Delta\theta}^{\frac{\pi}{2}+\Delta\theta} r^3 \sin{(\theta)} \, d\theta
\left[-T_{r\phi}{\bf\theta}_C+T_{r\theta}{\bf\phi}_C)
\right]^{r+\frac{\delta r}{2}}_{r-\frac{\delta r}{2}}.
\label{visc}
\end{eqnarray}
Thus the $z$-component of the angular momentum vector can only be
changed by the $T_{r\phi}$ component of the viscous stress tensor,
while the $x$ and $y$ components can also be changed due to
radial shear of vertical motion {\it via} $T_{r \theta}$.
The last term in Eq.~(\ref{torques}) which induces a 
change of the angular momentum vector of
disc material contained in a spherical shell is
due to the gravitational interaction with the secondary star:
\begin{eqnarray}
{\bf\dot{L}}^S(r)={\bf\dot{L}}^T(r)+{\bf\dot{L}}^A(r),
\label{second}
\end{eqnarray}
where the angular momentum change due to torques is given by:
\begin{eqnarray}
{\bf\dot{L}}^T(r)= \int_0^{2\pi}d\phi \int_{\frac{\pi}{2}-\Delta\theta}^{\frac{\pi}{2}+\Delta \theta}
{\bf t} \cdot r^2 \sin{(\theta)} \, d\theta \, \delta r
\end{eqnarray}
and the torque density ${\bf t}$ is given by:
\begin{eqnarray}
{\bf t}=-\rho{\bf r}\times{\bf\nabla}\Psi=\rho\left[\frac{1}{\sin{(\theta)}}
\frac{\partial\Psi}{\partial\phi}{\bf\theta}_C-
\frac{\partial\Psi}{\partial\theta}{\bf\phi}_C\right].
\end{eqnarray}
When working in the precessing frame there is an additional term
causing angular momentum change due to Coriolis and centrifugal forces:
\begin{eqnarray}
{\bf\dot{L}}^A(r)=\int_0^{2\pi}d\phi
\int_{\frac{\pi}{2}-\Delta\theta}^{\frac{\pi}{2}+\Delta \theta}
\rho \, r^2 \sin{(\theta)} \, d \theta \left(-r\Delta v_\phi{\bf\theta}_C+r\Delta v_\theta{\bf\phi}_C\right) \delta r \nonumber
\end{eqnarray}
with
\begin{eqnarray}
\Delta v_\phi=\left(-2{\bf\Omega}\times{\bf v}-{\bf\Omega}\times({\bf\Omega}\times{\bf r})\right){\bf\phi}_C\nonumber\\
\Delta v_\theta=\left(-2{\bf\Omega}\times{\bf v}-{\bf\Omega}\times({\bf\Omega}\times{\bf r})\right){\bf\theta}_C.
\end{eqnarray}
In the following we are only interested in the torque components
expressed in precessing frame coordinates.
Thus all the torque components have been calculated now for
simulations performed in the precessing frame.
However, when working in the binary frame,
we perform the above calculations and then project each of the
torque components onto precessing frame coordinate axes to ensure
a uniform approach to monitoring the torque contributions.
Since the latter are time dependent,
 the additional term when working in the binary frame is given by:
\begin{eqnarray}
{\bf\dot{L}}^A(r)=-{\bf\Omega}\times{\bf\hat{L}}\nonumber
\end{eqnarray}
This term accounts for the angular momentum change measured in
the precessing frame that arise because of the precession
of the frame.

\subsection{Calculation of precession and inclination rates}
We can now relate the above torques to the rate of change of
the precession and inclination angles.
When working in the precessing frame, the inclination angle $\delta$
and precession angle $\beta$ for disc material in a given radial shell,
as defined by Eqs.~(\ref{inclination}) and (\ref{precession}),
are given by:
\begin{eqnarray}
\cos(\beta)&=&\frac{\cos(\Omega_Ft)(\sin(\gamma_F)L_Z+\cos(\gamma_F)L_Y)+
\sin(\Omega_Ft)L_X}{\left[\left(\sin(\gamma_F)L_Z+\cos(\gamma_F)L_Y\right)^2
+L_X^2\right]^\frac{1}{2}}\nonumber\\
\cos(\delta)&=&\frac{-\sin(\gamma_F)L_Y+\cos(\gamma_F)L_Z}{|{\bf L}|}.
\label{angles}
\end{eqnarray}
Note that the vector components $L_X$, $L_Y$ and $L_Z$ in the above
expression, and in the remaining expressions in this subsection, should
have hat symbols because they are measured in the precessing frame.
We neglect these, however, in order to simplify the notation.
For sufficiently rigidly precessing discs the deviation from the mean
precession frequency $\Omega_F$ will be small, and the disc midplane
will remain close to the equatorial plane of the computational domain.
In this case we can use the approximation:
\begin{eqnarray}
L_X\ll L_Z\hspace{0.5cm}L_Y\ll L_Z.
\label{approx}
\end{eqnarray}
With this approximation, Eq.~(\ref{angles}) gives to first order:
\begin{eqnarray}
\delta&=&\gamma_F+\frac{L_Y}{L_Z}\nonumber\\
\beta&=&\Omega_Ft-\frac{1}{\sin{(\gamma_F)}}\frac{L_X}{L_Z}.
\label{approxangles}
\end{eqnarray}
Thus for sufficiently rigidly precessing discs the total inclination
is the sum of the inclination of the precessing frame and a small
deviation term which is proportional to the $y$-component of the
angular momentum vector, measured in the precessing frame.
Likewise the precession angle is the sum of precession due to the
frame and a small deviation term that is proportional to the $x$-component
of the angular momentum vector.
The advantage of this approximation is that the precession
and inclination angles become linear functions of the torque components.
This allows us to write:
\begin{eqnarray}
\frac{\partial\delta}{\partial t}&=&\left(\frac{\partial\delta}{\partial t}\right)^F+\left(\frac{\partial\delta}{\partial t}\right)^V+\left(\frac{\partial\delta}{\partial t}\right)^S\nonumber\\
\frac{\partial\beta}{\partial t}&=&\Omega_F+\left(\frac{\partial\beta}{\partial t}\right)^F+\left(\frac{\partial\beta}{\partial t}\right)^V+\left(\frac{\partial\beta}{\partial t}\right)^S
\label{totrate}
\end{eqnarray}
where each of the rates are calculated according to Eq.~(\ref{approxangles}):
\begin{eqnarray}
\left(\frac{\partial\delta}{\partial t}\right)^K&=&\frac{1}{L_Z}\dot{L_Y^K}-\frac{L_Y}{L_Z^2}\dot{L_Z^K}\nonumber\\
\left(\frac{\partial\beta}{\partial t}\right)^K&=&-
\frac{1}{\sin(\gamma_F)}\left(\frac{1}{L_Z}\dot{L_X^K}-
\frac{L_X}{L_Z^2}\dot{L_Z^K}\right). \label{rates}
\end{eqnarray}
Hence, in circumstances where the approximation expressed in
Eq.~(\ref{approx}) holds, we can measure separate contributions to
the inclination and precession rate coming from the radial flux
(Eq.~\ref{flux}), viscous stress (Eq.~\ref{visc}) and gravitational
interaction with the secondary star (Eq.~\ref{second}).
When discussing contributions to differential precession
rates later in the paper, we will omit the constant mean precession rate
$\Omega_F$ of the precessing frame in the expression for the total rate
given by Eq.~(\ref{totrate}).

%We use 452 grid cells in radial and 600 cells in azimuthal direction.
%In meridional direction the extend was varied depending on the initial tilt angle and pressure scale height of the disc model, but the resolution stayed fixed to 5 grid cells per pressure scale height.

\section{Linear theory of free warps and bending waves}
In order to calibrate the code, we now examine how well it is
able to model the propagation of free bending waves, and
compare the results with linear theory.
The linear theory of warps in accretion discs was initially investigated
by \cite{pringle}.
Their analysis is valid in a regime in which the dimensionless
viscosity parameter $\alpha$ \citep{shakura} lies in the
range $h\le\alpha\le 1$. In this regime, globally warped discs were
found to evolve diffusively, with a diffusion coefficient larger
than that associated with mass flow through the disc by a factor of
$\sim\frac{1}{2\alpha^2}$, assuming an isotropic viscosity.
\cite{paplin} investigated the regime in which  $\alpha\leq h$,
and showed under that, for low frequency modes, the governing equation for
the disc tilt changes from being a diffusion equation to a wave equation,
such that warps are communicated through the disc via propagation
of bending waves. In a  keplerian disc these are expected
to propagate with little dispersion at half the sound speed, $c_s/2$.

\begin{figure*}
\begin{center}
%\resizebox{10cm}{8.86cm}{\includegraphics[width=100mm,angle=0]{./one.jpg}}
\includegraphics[width=100mm,angle=0]{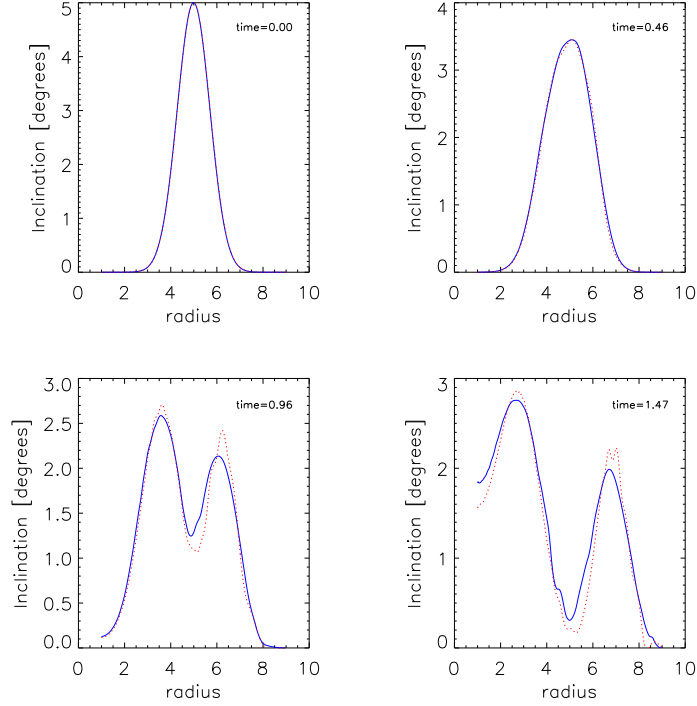}
\end{center}
\caption{Time evolution of free warp for h=0.03, $\delta_{MAX}=5^\circ$.
NIRVANA results (blue solid line); linear results (red dashed line).}
\label{one}
\end{figure*}

\citet{nelsonpap} used smooth particle hydrodynamic simulations to examine
the propagation of free bending waves, comparing their numerical results
with linear calculations. Overall, they found that the SPH code managed
to model the propagation of bending waves quite accurately as long as
the warp amplitude remains small such that linear theory holds.
One drawback associated with using SPH for that problem, however,
was the fairly large numerical viscosity associated with the scheme
which is difficult to control and specify {\it ab initio}.

In this section we compare our numerical results with a linear calculation,
adopting a similar set-up to the one used by \citet{nelsonpap}.
The equations describing the evolution of linear bending waves
have been derived by \cite{nelsonpap}. We do not reproduce the
full derivation here, but rather mention some of the salient points
associated with the derivation.

\subsection{An initial value problem}
Small-amplitude warps can be considered to be linear perturbations of
a disc with midplane initially coincident with the (${\bf e}_1$, ${\bf e}_2$)
plane. Taking the $\phi$ dependence of the perturbations to be
$\propto e^{im\phi}$, with $m=1$ for global warps, the Euler equations written
in cylindrical coordinates $(r,\phi,z)$ can be linearised \citep{paplin,nelsonpap}.
An analytical solution to the linearised equations corresponds to
the case of a rigid tilt. In this case the velocity and pressure
perturbations are given by:
\begin{eqnarray}
&v_r'&=-z\Omega_K \delta\nonumber\\
&v_\phi'&=-iz\delta\frac{d(r\Omega_K)}{dr}\nonumber\\
&v_z'&=r\Omega_K \delta\nonumber\\
&W'&=\rho'\frac{c_S^2}{\rho}=-irz\Omega^2_K\delta
\label{linsolution}
\end{eqnarray}
where $\Omega_K$ is the Keplerian angular frequency and
$\delta$ is a small constant inclination.
For large scale warps, it is expected that the local
inclination $\delta$ varies on a length scale significantly
larger than the disc thickness $H=h\cdot r$. It is then reasonable
to assume the inclination is also approximately independent of $z$.
Using these assumptions, and the rigid tilt solution (\ref{linsolution}),
the $z$ dependence of the linearised equations disappears
when introducing the new quantities $q_r'=\frac{v_r'}{z}$ and
$q_\phi'=\frac{v_\phi'}{z}$.
The resulting expressions are given by \cite{nelsonpap} in the form:
\begin{eqnarray}
\frac{\partial q_r'}{\partial t}&+&i\Omega_K q_r'-2\Omega_K q_\phi'=i\frac{\partial(r\Omega^2_K\delta)}{\partial r}\nonumber\\
\frac{\partial q_\phi'}{\partial t}&+&i\Omega_K q_\phi'+q_r'\frac{1}{r}\frac{\partial(r^2\Omega_K)}{\partial r}=-\Omega^2_K\delta\nonumber\\
\frac{\partial v_z'}{\partial t}&+&i\Omega_K v_z'=ir\Omega^2_K\delta\nonumber\\
{\cal F} r&\Omega^2_K&\left(\frac{\partial\delta}{\partial t}+i\Omega_K\delta\right)=-\frac{i}{r}\frac{\partial (r\mu q_r')}{\partial r}+\frac{\mu q_\phi'}{r}+i\Sigma v_z'
\label{linearized}
\end{eqnarray}
where
\begin{eqnarray}
&{\cal F}&=\int_{-\infty}^\infty\frac{\rho z^2}{c_S^2} dz\nonumber\\
&\mu&=\int_{-\infty}^\infty\rho z^2 dz\nonumber
\label{expressions}
\end{eqnarray}
Equations (\ref{linearized}) are a set of coupled differential
equations which describe the evolution of the inclination $\delta$
as a function of $r$ and $t$. Thus they can be used to follow the
evolution of a bending disturbance from initial data.
Below we use the time-dependent solution calculated in this way to
test NIRVANA's ability to propagate bending disturbances.

Assuming a time dependence of the perturbations
$\propto e^{i\sigma t}$, with the wave frequency obeying the relation
$\sigma\ll\Omega_K$ for slowly varying warps, one can derive algebraic equations
for the perturbed quantities. These show that warps induce horizontal motions
and vertical shear in the disc through terms proportional to
$\left( \frac{\partial v_r'}{\partial z} \right.$,
$\left. \frac{\partial v_\phi'}{\partial z}\right)$ \citep{paplin}.
Physically, these horizontal motions are generated by pressure forces
in the disc which arise from the vertical misalignment of neighboring
regions of the disc midplane. They are responsible for driving
the bending wave, which propagates with approximately half the
sound speed.

\begin{figure*}
\begin{center}
\includegraphics[width=100mm,angle=0]{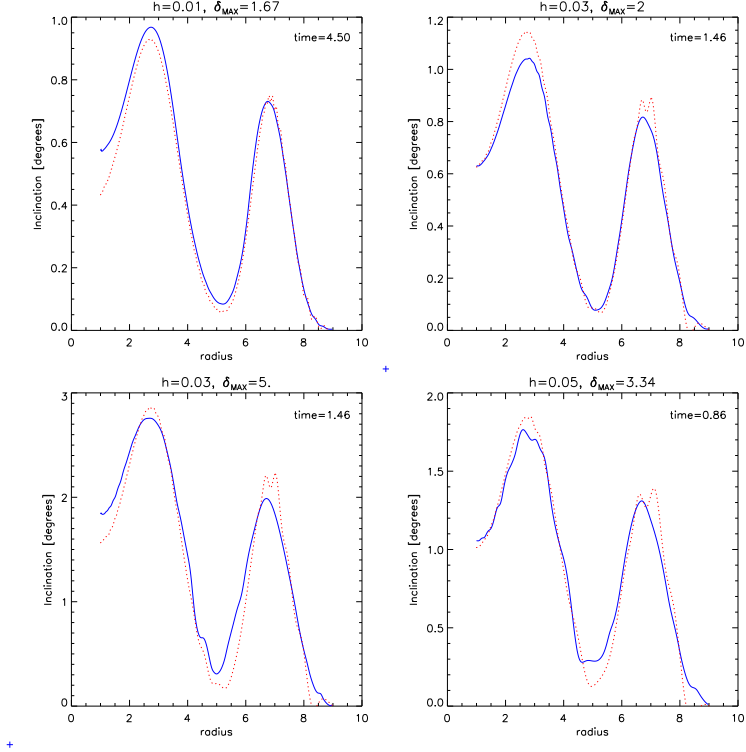}
\end{center}
\caption{Final stage of evolution for different sets of disc parameters.
NIRVANA solution is shown by the solid blue line, the
linear calculation by the red dashed line.}
\label{all}
\end{figure*}

The main effect of viscosity in the disc is to damp these horizontal
motions, reducing the efficacy of bending wave propagation.
As the viscosity increases and $\alpha > h$, the bending disturbance
begins to evolve diffusively on a time scale that scales inversely with
$\nu/(2\alpha^2)$ \citep{pringle}, where this latter quantity is now
the diffusion coefficient associated with warp propagation in this
regime.
If the viscosity is small enough that its effect on the
unperturbed quantities can be neglected, then
the set of equations (\ref{linearized}) can be extended to include the
effect of a small viscosity by the replacement \citep{paplin}:
\begin{eqnarray}
\left(\frac{\partial}{\partial t}+i\Omega_K\right)\rightarrow
\left(\frac{\partial}{\partial t}+(i+\alpha)\Omega_K\right).
\label{opreplacement}
\end{eqnarray}
If the viscosity is large, however, then the complete viscous stress
tensor should be included in the equations of motion, making the analysis
rather complicated, since there would then
be accretion velocities in the unperturbed state.

\subsection{Setting up initial data}
A number of simulations using NIRVANA have been performed,
in which warps with different amplitudes in discs with different
parameters were set up and allowed to evolve.
We compare these solutions with those obtained using the linear theory
expressed in Eqs.~(\ref{linearized}) using the same set of initial conditions.
The linear calculations employed a one dimensional finite
difference scheme for which we used 1000 equally spaced
grid cells. Starting with a gaussian initial inclination profile:
\begin{eqnarray}
\delta(r)=\delta_{MAX}e^{(-|r-r_{MID}|^2)}\nonumber
\end{eqnarray}
with $r_{MID}=5$ to be centred in the middle of a disc with
inner radius $r=1$ and outer radius $r=9$,
the set of equations (\ref{linearized}) was integrated forward in time.
We chose the viscosity in these test simulations to be
$\alpha=10^{-3}\ll h$ since we are mainly interested in the
regime of propagating bending waves, and because the linear solution
presented here becomes invalid for large viscosities.
The NIRVANA code was set up in an inertial frame using
the expressions (\ref{density}) and (\ref{tiltvelocity})
to specify the initial density and velocity, in combination with
Eq.~(\ref{sintheta}), with $\gamma_F(r)=\delta(r)$ now being the
disc tilt which is a function of $r$.
In these simulations we used $N_r=404$ cell in radius,
5 cells per pressure scale height in the meridional direction
(and different numbers of vertical scale heights, depending on the
initial warp amplitude, as described below) and
$N_{\phi}=300$ cells in azimuth. We used the same boundary condition as for the global simulations, which were described in section 3.1.

\subsection{Results}
An example result is shown in Fig.\ref{one} which has a disc aspect
ratio of $h=0.03$ and maximum initial inclination angle $\delta_{MAX}=5^\circ$.
In this particular simulation the meridional domain contains 15
pressure scale heights ($\Delta \theta=12.89^\circ$) in total.
Fig.\ref{one} shows the inclination as a function of radius at
different stages of the evolution. The solid blue line represents the
NIRVANA solution, while the red dashed line shows the linear calculation.

Moving from the top left to top right panel, we can observe how the
initial pulse begins to broaden. At time $t=0.96$ we see that
the pulse has separated into an in-going and out-going bending wave.
By time $t=1.46$ the pulses have reached the boundaries of the domain
and have fully separated. It is clear that NIRVANA is able
to capture the evolution of these low amplitude bending waves with a
high level of accuracy.

In \cite{nelsonpap} the numerical SPH and linear solutions showed more
substantial disagreement at the position of the initial pulse ($r=5$)
at the final stage of the simulation, with the separation of the two
pulses not being captured as accurately. This was probably due to
the larger numerical diffusion present in the SPH code, though
it should be remembered that the SPH simulations were performed using
only 20,000 particles.
The more accurate solution obtained with NIRVANA, on the other hand,
is a clear indication that the numerical diffusion is smaller.

We also performed simulations for other values of the disc
thickness $h$ and maximum initial inclination angle $\delta_{MAX}$.
The number of pressure scale heights used in the meridional
direction was scaled with the maximum inclination angle to
avoid the disc interacting with the meridional boundaries.
The results are shown in Fig.\ref{all}, and are presented at a
stage of the simulations when the bending waves have propagated
all the way to the radial boundaries. This corresponds to the
final panel of Fig.\ref{one}. Since the bending waves propagate
with half the sound speed, the evolution occurs over a
longer time in the thinner discs ($h=0.01$), and over a shorter time
in the thicker disc models ($h=0.05$).
As can be seen from Fig.\ref{all}, the agreement between the
numerical solution and the linear calculation is good in all
the models presented here. This is particularly the case for
those runs with lower initial maximum inclinations in Fig.\ref{all}
(top left and right panels), but the results are still in very reasonable
agreement for the higher inclination cases in Fig.\ref{all} (lower
left and right panels).

We mention here that for even higher inclination angles
NIRVANA behaves more diffusively, and the agreement between the linear
and non linear solutions gets worse.
As found by \citep{nelsonpap}, when the warps become non-linear they
generate horizontal motions that are of the order of the sound speed, creating
shocks because of the symmetric form of the initial bending
disturbance. The result is an effectively higher viscosity operating in
the disc, which is clearly not accounted for in the linear calculations.

\section{Tilted discs in binary star systems}

\begin{table*}
\begin{center}
\begin{tabular}{cccccccccc}
\hline
\hline
Model& $h$ & $\alpha$ & $\gamma_F$ & $R$ & Frame& $\tau_W$ & $\tau_P$ & Predicted & Observed\\
label&     &          &            &     &      &          &          & behaviour & behaviour\\
\hline
1&0.05&0.025&45&9&precessing&5.43&48.01&rigid precession & rigid precession\\
2&0.05&0.1&45&9&precessing&10.87&47.60&rigid precession& rigid precession\\
3&0.03&0.015&45&9&precessing&9.05&43.90&rigid precession& rigid precession\\
4&0.03&0.1&45&9&precessing&30.19&47.02&rigid precession& rigid precession\\
5&0.03&0.1&25&9&precessing&30.19&36.37&rigid precession& rigid precession\\
6&0.01&0.005&10&8&binary&22.77&33.87&rigid precession& rigid precession\\
7&0.01&0.1&10&8&binary&227.7&36.25&disrupted/broken& rigid precession\\
6a&0.01&0.005&10&10&binary&31.83&21.48&disrupted/broken& broken\\
7a&0.01&0.1&10&10&binary&318.32&19.12&disrupted/broken& disrupted \\
\hline
\end{tabular}
\end{center}
\caption{Table of runs. The disc is characterized by the aspect ratio, $h$,
viscosity parameter, $\alpha$, and initial inclination angle $\gamma_F$.
We also list the initial disc outer radius $R$ and the reference frame in
which the simulation was performed. The warp propagation times, $\tau_W$ and
differential precession times, $\tau_P$, for each of the models are also listed
and indicate the theoretically expected and observed behaviours.}
\end{table*}

We now discuss our results for the misaligned, tidally interacting discs.
We set up the disc models according to the procedure outlined in Sect. 3,
and details of the model parameters can be found in Table 1.
The models are characterized by the disc aspect ratio, $h$,
viscosity, $\alpha$, and initial inclination with respect to the
binary orbit plane, $\gamma_F$. The disc outer edge, $R$,
was chosen such that the disc is already very close 
to its tidal truncation radius,
which is about one third of the distance between secondary and primary
star, $D$. In some models the initial radius of the outer disc edge
was varied also. 

In this study, we are interested in examining the
evolution of discs that fulfil the condition for wave-like warp propagation,
$h > \alpha$, as well as discs that support diffusive warp propagation,
$h < \alpha$. For the models in the wave regime we set
$\alpha=\frac{1}{2}h$, and for the runs in the diffusive regime we set
$\alpha=0.1\gg h$.

%\begin{table}[b]
%\begin{center}
%\begin{tabular}{ccccc}
%\hline
%\hline
%Model& $\tau_W$ & $\tau_P$ & Predicted & Observed\\
%label&  &  & behaviour & behaviour\\
%\hline
%1&5.43&48.01&rigid precession & rigid precession\\
%2&10.87&47.60&rigid precession& rigid precession\\
%3&9.05&43.90&rigid precession& rigid precession\\
%4&30.19&47.02&rigid precession& rigid precession\\
%5&30.19&36.37&rigid precession& rigid precession\\
%6&22.77&33.87&rigid precession& rigid precession\\
%7&227.7&36.25&disrupted/broken& rigid precession\\
%6a&31.83&21.48&disrupted/broken& broken\\
%7a&318.32&19.12&disrupted/broken& disrupted \\
%\hline
%\end{tabular}
%\end{center}
%\caption{This table lists the warp propagation times, $\tau_W$ and
%differential precession times, $\tau_P$, for each of the models,
%and indicates the theoretically expected and observed behaviours.}
%\end{table}

The perturber is evolved on a circular orbit at a distance of
$D=30$, and its mass is increased linearly from zero up to
$M_S=M_P$ over a time interval of 4 orbits, during which time its
distance is kept constant.
Models 1 -- 5 were performed in the precessing frame.
If we use Eq.~(\ref{omega}) to estimate the precession frequency, then
we obtain a value of $\Omega_F=-4.86 \cdot \cos(\gamma_F)$ [degrees/orbit].
We find, however, that a better fit to test simulations is given
by $\Omega_F=-3.74\cdot cos(\gamma_F)$ [degrees/orbit], which is the value
used in the simulations. The reason is that as the disc gets tidally
truncated it tends to precess slower, as can be seen from Eq.~(\ref{omega}).

The resolution was chosen such there are 5 cells per pressure scale height
in the meridional direction for all models. Models 1 and 3 incorporated
15 scale heights in the meridional direction. Models 2, 4 and 5
used 22.5 scale heights (to allow for a greater degree of twisting
in the higher viscosity models where warp communication is expected
to be less efficient).
In the radial and azimuthal directions we used $N_r=404$ and $N_{\phi}=300$
cells, respectively, for each of these models.
\begin{figure*}
\begin{center}
\resizebox{15.12cm}{14.72cm}{\includegraphics[width=100mm,angle=0]{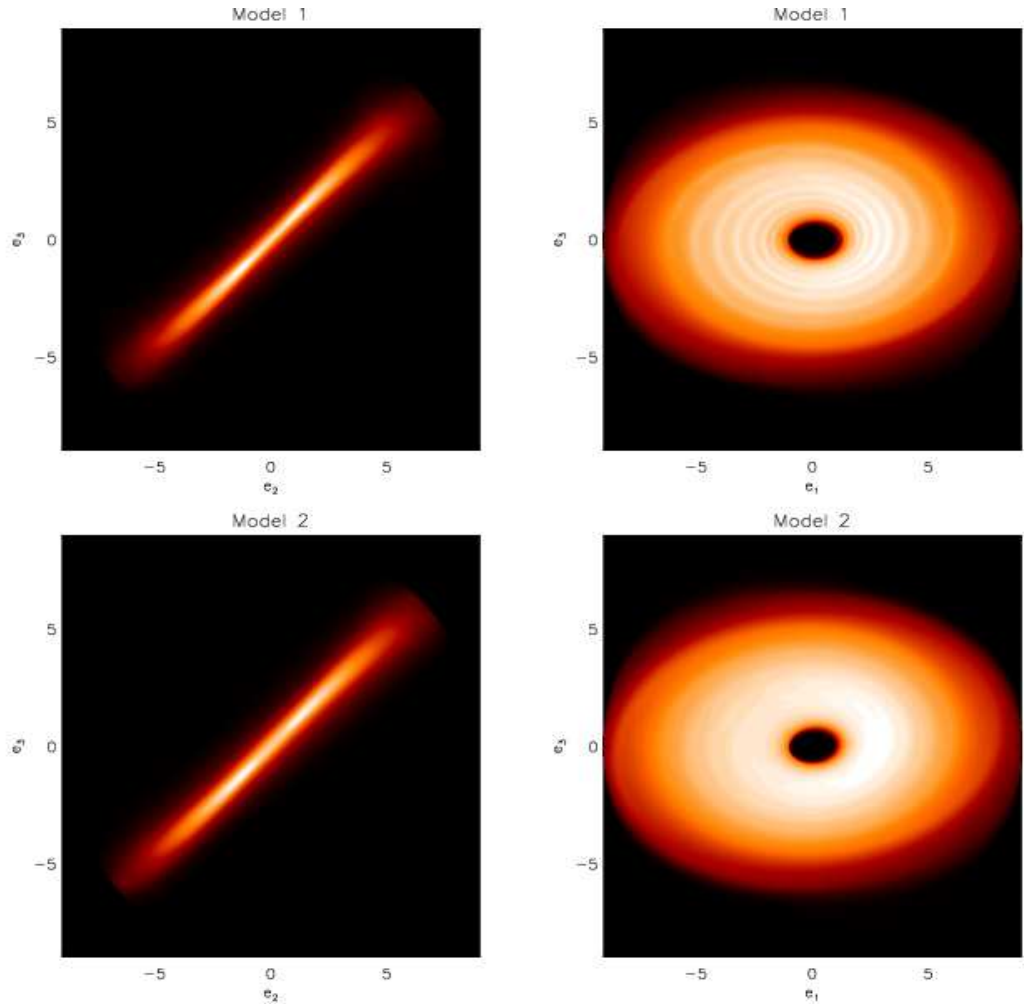}}
\end{center}
\caption{Column density plot at time $t=75.0$ for Model 1 (upper panels),
and Model 2 (lower panels). The left panels correspond to projection onto
the (${\bf e}_2$, ${\bf e}_3$) binary plane, and the right panels to projection
onto the (${\bf e}_1$, ${\bf e}_3$) plane.}
\label{images52}
\end{figure*}

Models 6 and 7 were performed in the non-precessing binary frame.
Since the disc midplane is inclined with respect to the equatorial plane of
the computational grid in these simulations, higher resolutions are
necessary to avoid numerical errors. This is particularly true in the
azimuthal direction because tilting the disc causes a component of the
disc vertical structure to point along this direction.
For a disc whose midplane coincides with the equatorial plane of the 
computational grid, pairwise cancelation of fluxes ensures 
conservation of angular momentum to machine accuracy. However, 
if the disc midplane is inclined with respect to the equatorial 
plane of the computational grid, the accuracy is limited by the 
advection scheme, and higher resolution in the azimuthal direction 
becomes necessary to avoid unphysical evolution of precession 
and inclination angles.
Hence we used 600 cells in azimuth.
In order to be able to resolve spiral waves for these 
extremely thin discs, we used 1056 cells
in radius. The disc outer edge was chosen to be a bit smaller
than in the thick disc runs, since highly pronounced non linear effects at the
outer edge were seen in lower resolution test simulations (described as
Models 6a and 7a in Table 1). In effect a strongly
perturbed and narrow outer rim of gas became partially detached
from the main body of the disc and developed a very different
inclination and precession angle evolution in Model 6a, an effect
not observed in Model 6 with a smaller initial outer radius. 
This phenomenon is discussed in more detail later in the paper. 
In order to accommodate an inclined disc with an inclination of
$10^\circ$ we used 60 pressure scale heights ($\sim\pm 17.03$ degrees)
in the meridional direction, with 5 cells per pressure scale height.

We now describe the results of these models, where we have ordered
our discussion in terms of the disc aspect ratio, $h$.

\begin{figure*}
\begin{center}
\resizebox{18cm}{10.5cm}{\includegraphics[width=100mm,angle=0]{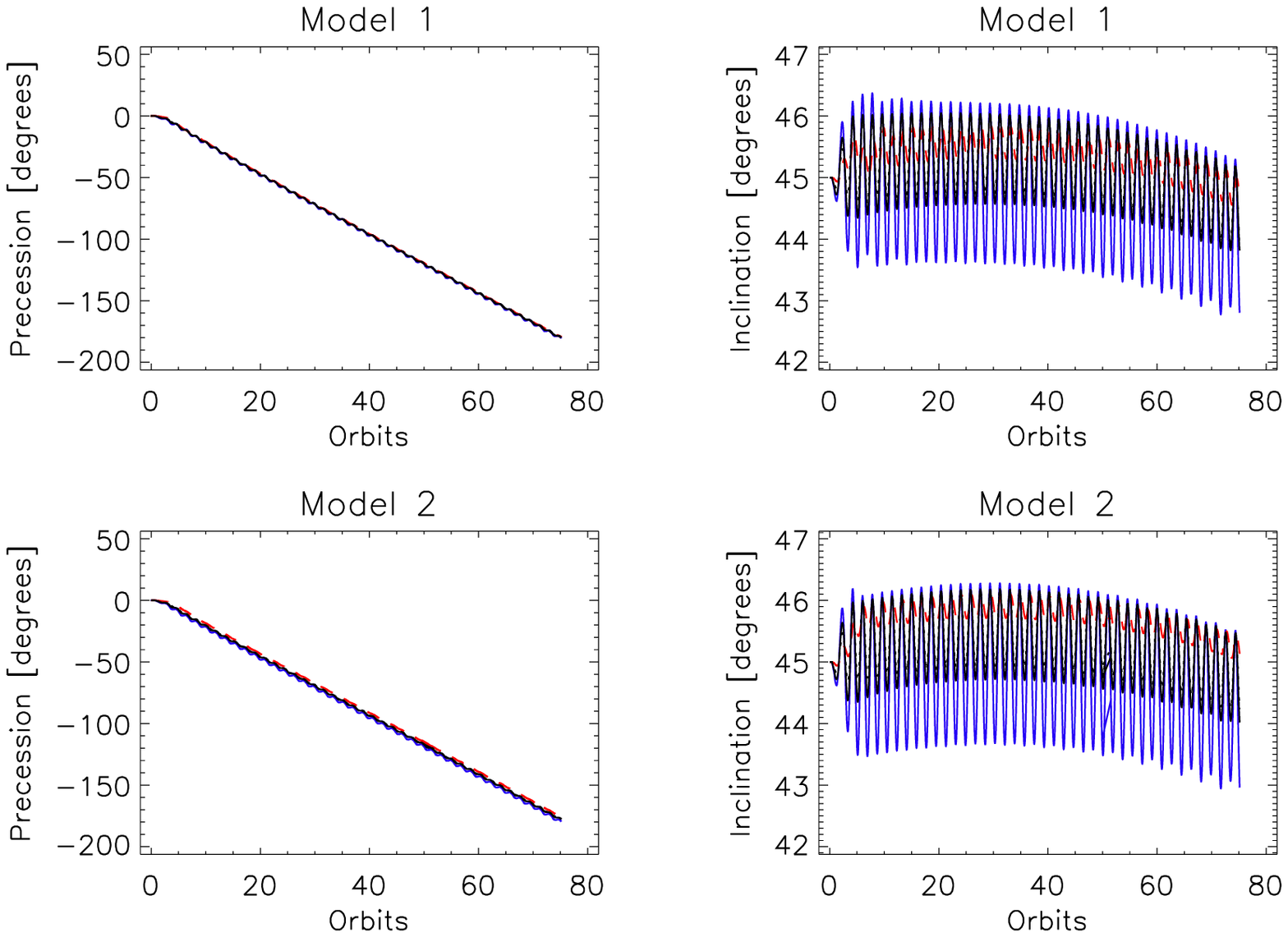}}
\end{center}
\caption{Evolution of precession (left panels) and inclination angles
(right panels) for Models 1 and 2.
The blue solid line corresponds to disc material
between $r=8$ and $r=9$, the red dashed line to disc material between
$r=1$ and $r=2$, and the black dotted line to disc material between
$r=4$ and $r=5$. The black solid line represents the entire disc.}
\label{full52}
\end{figure*}

\subsection{Models 1 and 2}
A column density plot for the disc in Model 1 is displayed in 
the upper panels of Fig.\ref{images52} 
corresponding to the end stage of this simulation.
The left panel displays a projection of the disc column density
onto the (${\bf e}_2$, ${\bf e}_3$) plane of the binary reference frame,
and the right panel displays a projection onto the binary
(${\bf e}_1$, ${\bf e}_3$) plane. At this stage the 
disc has precessed rigidly to $\beta=-180$ degrees,
so the disc appears edge on in the (${\bf e}_2$, ${\bf e}_3$)
plane and face on in the (${\bf e}_1$, ${\bf e}_3$) plane. 
The precession of 180 degrees is evident,
as the disc angular momentum vector now has a
negative ${\bf e}_2$ component, whereas it had a positive ${\bf e}_2$ component
in its initial state, as can be seen from the transformation given
by Eq.~(\ref{transform}) at $t=0$.
Another result of the interaction with the secondary star is the
formation of spiral waves, which are apparent in Fig.\ref{images52}.

Fig.\ref{full52} shows the evolution of the precession ($\beta$) and inclination ($\delta$) angles,
respectively, for disc material orbiting at different radii.
These have been calculated using Eqs.~(\ref{precession}) and (\ref{inclination}) respectively, where ${\bf L}$ has been calculated using Eq.~(\ref{angmom}), which we have
integrated over radial shells of width $\Delta r=1$. 
Thus the red dashed line in Fig.\ref{full52} corresponds to disc gas
between $r=1$ and $r=2$, the blue solid line to disc gas between $r=8$ and $r=9$,
and the black dotted line to disc gas between $r=4$ and $r=5$. 
The solid black line shows the precession and inclination 
angles of the entire disc.
It is apparent that the lines are almost indistinguishable in
Fig.\ref{full52} (upper left panel), showing that the different disc
parts in Model 1 precess at the same uniform rate, such that the disc is
in a state of rigid body precession with almost no twist.
This behaviour is expected from linear theory \citep{papterquem}. 
In Table 1 we show the warp propagation timescale $\tau_W$ and the
differential precession timescale, $\tau_P$. 
In the bending wave propagation regime ($h>\alpha$), $\tau_W=2R/c_S$, 
since warps propagate with half the sound speed in this regime. 
In the diffusive regime ($h<\alpha$) $\tau_W=2\alpha^2R^2/\nu$. 
The warp propagation timescale should be compared to the 
differential precession timescale $\tau_P\sim 2/\Omega_F$.
If $\tau_W<\tau_P$ it is expected that the disc will precess as a 
rigid body, while if $\tau_W>\tau_P$ the disc may
become disrupted as a result of differential precession.
Hence, we see that the rigid precession of the disc in Model 1 agrees
with expectations.
Our Model 1 is very similar to Model 1 in \cite{larwood}, which also shows
rigid body precession.
The inferred precession period in our Model 1 is
about 146 orbits, which is a little larger than the period given by
Eq.~(\ref{omega}). The difference arises because the disc is
tidally truncated and shrinks slightly, so its precession rate decreases.

Looking at the upper right panel of Fig.\ref{full52}, we observe that the
inclination for Model 1 changes only slightly during the simulation.
The oscillation seen in the inclination rates is driven by the secondary star.
Since the orbital plane is inclined with respect to the disc midplane,
the vertical component of the gravitational force due to the secondary star
causes the disc to perform a forced oscillation with a frequency equal
to twice the binary frequency.
\begin{figure*}
\begin{center}
\resizebox{15.12cm}{14.72cm}{\includegraphics[width=100mm,angle=0]{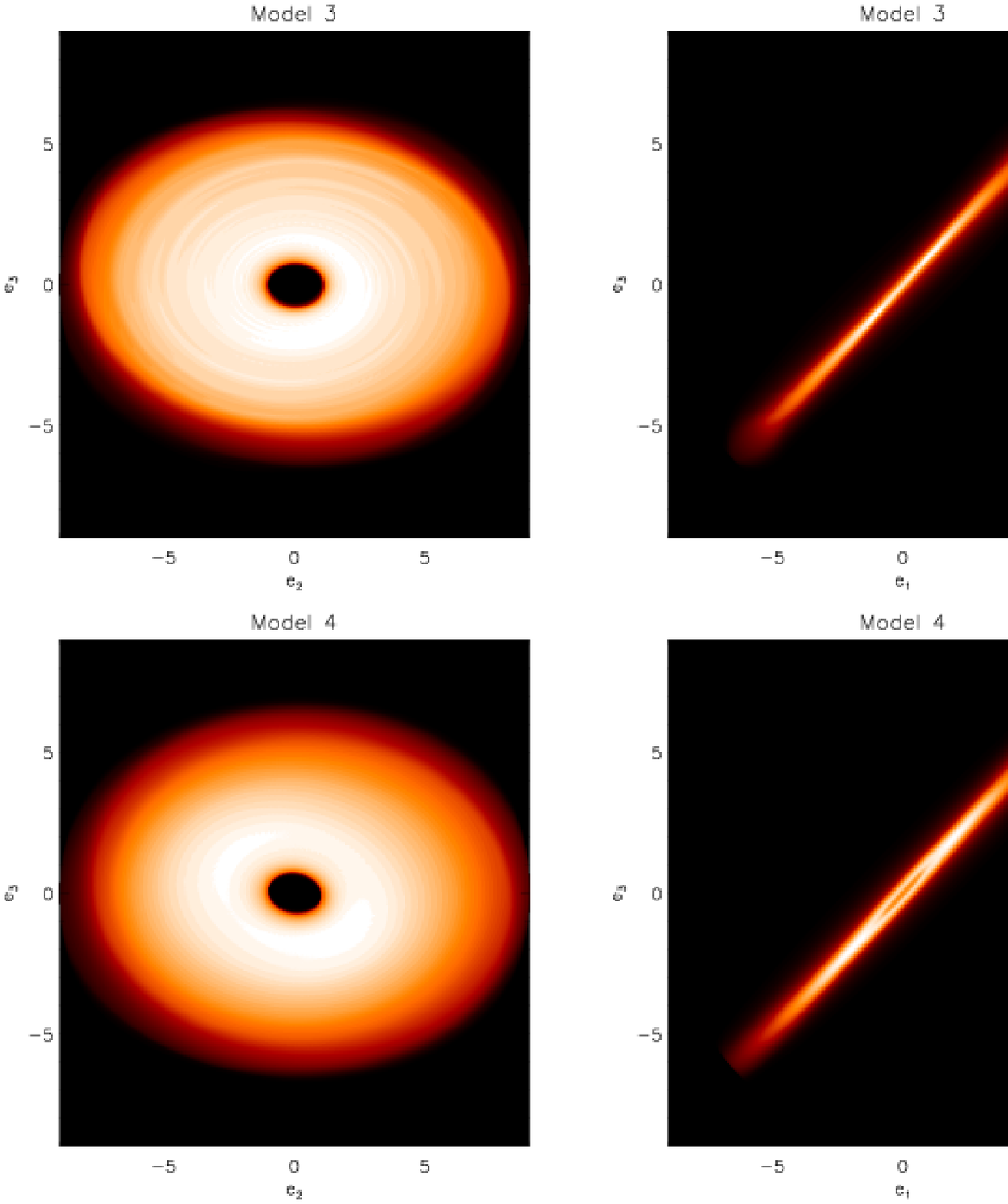}}
\resizebox{15.12cm}{4.57cm}{\includegraphics[width=100mm,angle=0]{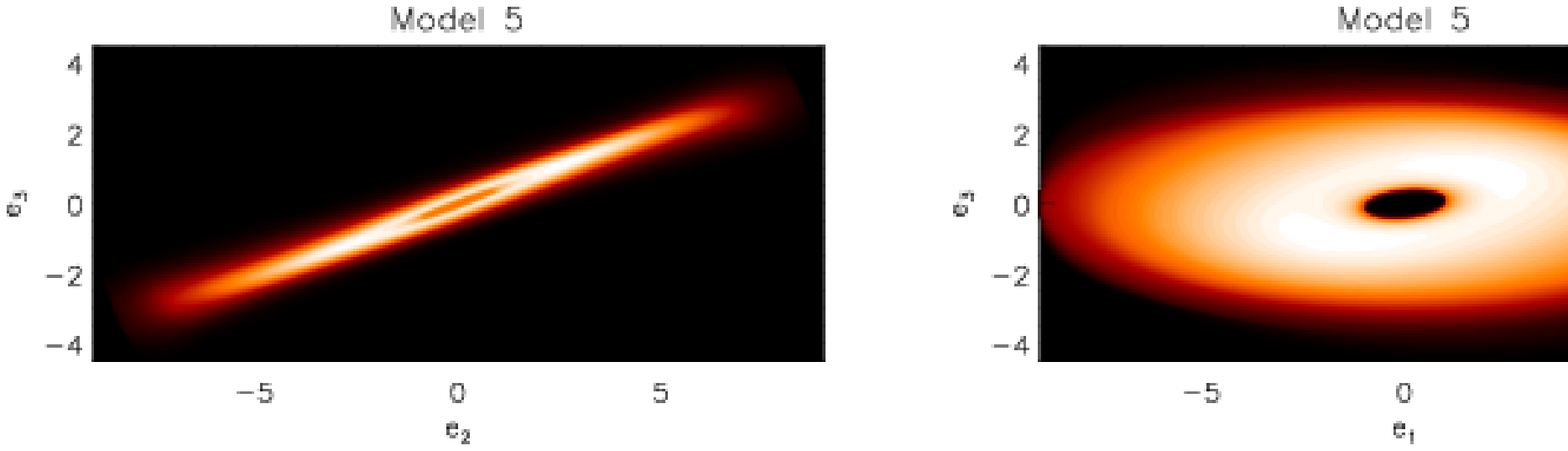}}
\end{center}
\caption{Column density plots for Model 3 (upper panels) and
 Model 4 (middle panels) at time $t=35.0$,
 and Model 5 at time $t=55.0$.
The left panels correspond to projection onto the
(${\bf e}_2$, ${\bf e}_3$) plane, the right panel to projection onto
the (${\bf e}_1$, ${\bf e}_3$) plane.}
\label{images51}
\end{figure*}
\cite{papterquem} argued that the disc is expected to align with
the binary orbit on a time scale similar to that over which
binary torques cause the total angular momentum of the disc to evolve.
If the viscously-induced outward expansion of the disc is balanced
by tides due to the companion, then disc-alignmnent should occur on
the viscous evolution time, $T_\nu$. For Model 1 we estimate 
$T_\nu=R^2/\nu \sim 2175$ orbits. Extrapolating the disc evolution shown in
Fig.~\ref{full52} gives an alignment time of $\sim 2025$ orbits, in good agreement
with the viscous time scale.

\citet{lubow} present images which represent disc 
shapes for a variety of disc parameters, obtained using a
linear analysis of tilted discs in binary systems.
They introduce a parameter $\epsilon$, which is a measure of
the disc thickness, and is roughly equivalent to $\sqrt{6} H/R$. 
Their model a displayed in their Figure 7 has $H/R \simeq 0.041$, 
$\alpha=0.01$ and $D/R=0.3$, similar to our model 1.
Although a detailed comparison is not possible, it appears
that there is general agreement between linear theory and
our non-linear simulation, since both show almost no distortion of
the disc due to twisting or warping.

Column density plots for Model 2 are displayed in
the lower panels of Fig.\ref{images52}.
Warp propagation in this higher viscosity model ($\alpha=0.1$)
is expected to be diffusive, and therefore less efficient
than for Model 1 (as shown by the warp propagation times listed in Table 1).
This run was also evolved
until the precession angle reached -180 degrees. The high 
viscosity operating in the disc leads to the damping of the spiral
waves before they can propagate very far, and thus they are not apparent
in the images.

The precession angles for Model 2 are displayed
in Fig.\ref{full52} (lower left panel).
We observe that this more viscous disc develops a small differential twist
before it attains a state of rigid precession. This suggests that
the disc needs to develop a slightly more distorted shape in order for
internal stresses to counterbalance the companion-induced differential
precession when warp communication is less efficient. Nonetheless, rigid
body precession is expected in this case, and the numerical results
are in agreement with this expectation.

Examining the inclination evolution for Model 2 shown in the lower right
panel of Fig.~\ref{full52}, we see it is very similar to that observed for
Model 1. The viscous time scale in Model 2 should be approximately 4 times
shorter than for Model 1, but there is no indication that the inclination
evolution is faster in this case. We note that the inclination evolution rates
do not appear to have reached final steady values for either Models 1 or 2
by the end of the simulations, such that an accurate assessment of the longer
term evolution rate is not possible. 
This may be because the outer disc radius has not had time to
equilibrate during the simulations as they have only run for
a fraction of the global viscous timescale of the disc.
We are able to conclude,
however, that the inclination evolution occurs over timescales
much longer than the precession time, and appears to be consistent
with evolution on the global viscous time of the disc.

\begin{figure*}
\begin{center}
\resizebox{18cm}{15.75cm}{\includegraphics[width=100mm,angle=0]{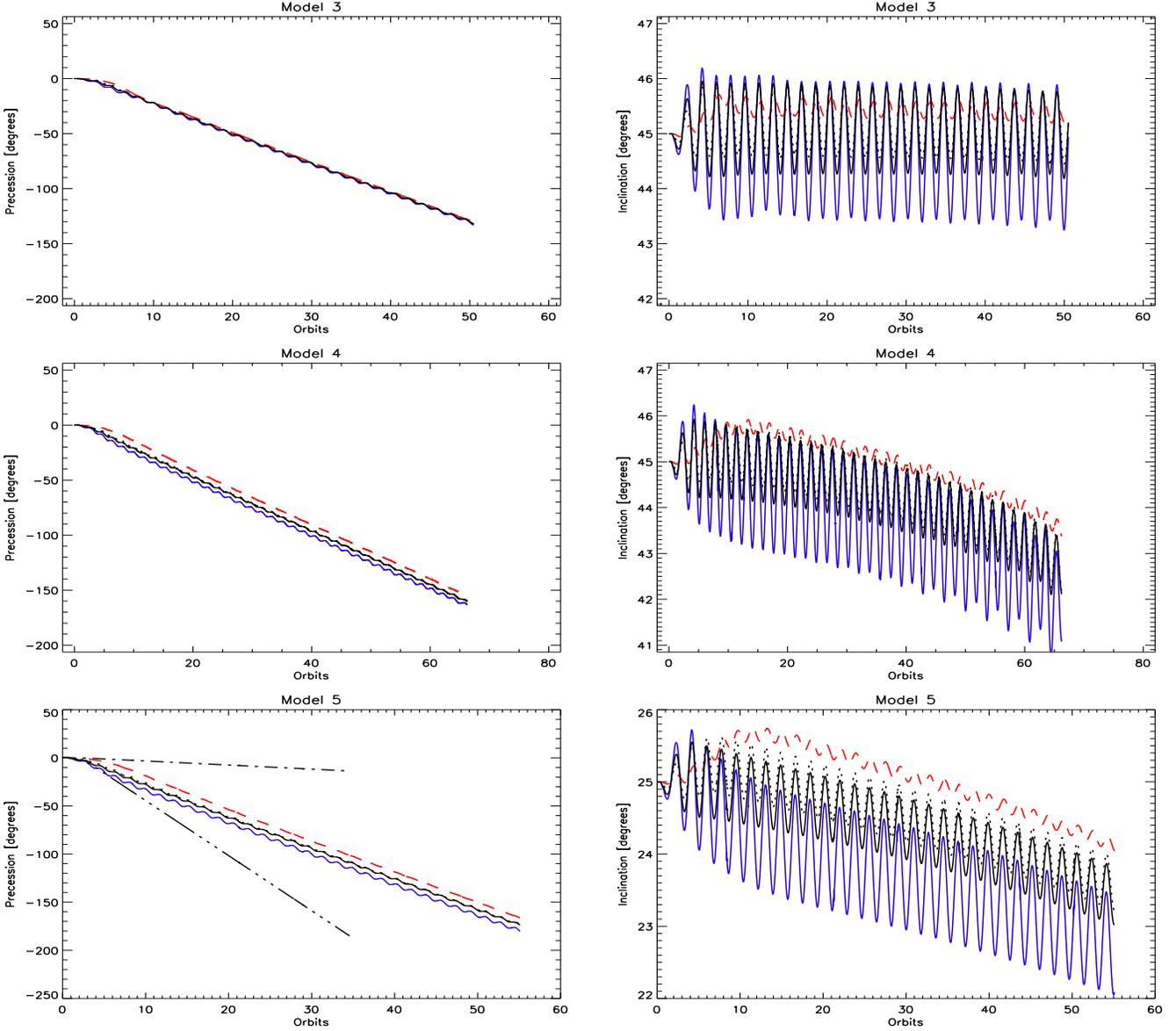}}
\end{center}
\caption{Evolution of precession (left panels) and inclination angles 
(right panels). The blue solid line corresponds to disc material between 
$r=8$ and $r=9$, the red dashed line to disc material between 
$r=1$ and $r=2$ and the black dotted line to disc material between 
$r=4$ and $r=5$. The solid black line represents the entire disc. 
In the lower left panel the free particle precession rates for the inner 
(dashed-dotted line) and outer disc (dashed-triple-dotted line) 
are also depicted for Model 5. These have been calculated using Eq.~(\ref{free}).} 
\label{full51}
\end{figure*}

\subsection{Models 3, 4 and 5}
As shown in Table 1, Models 3, 4 and 5 all have $h=0.03$.
Model 3 has $\alpha=0.015$, and so warp propagation
occurs {\it via} bending waves, whereas Models 4 and 5
have $\alpha=0.1$ such that warps propagate diffusively.
Table 1 also shows that the warp propagation for Models 4 and 5
is expected to be considerably less efficient than
for Model 3.
Models 3 and 4 had inclinations of $45^{\circ}$, whereas
Model 5 had an inclination of $25^{\circ}$, which induces a more
rapid precession of the disc, and therefore has a greater tendency to 
twist the disc up. 

%{\bf A feature observed after very long evolution times in our disc models which is outside the scope of this study, is the growth of eccentricity.
%The eccentricty growth timescale was shown to be very sensitive on the disc parameters \citep{kley-pap}. In particlular it was shown that the timescale becomes shorter for thinner and more viscous discs. In our disc models the disc was seen to become eccentric after about 60 orbits (16 binary orbits), which agrees well with growth timescales given by \citep{kley-pap} for an equal mass perturber and disc parameters close to the ones used in our study.
%Because the interaction of an eccentric disc with the open boundary conditions resulted in substantial mass lost form the system, we switched off the simulations, once the discs were found to become eccentric.
%Hence we were unable to evolve Models 3 and 4 until they
%had precessed by $180^{\circ}$. Therefore the upper and middle panels
%of Fig.~\ref{images51} display the column density images of the disc
%at times corresponding to the moment when the outer disc
%has precessed by $90^{\circ}$ instead.} 
At the outset of this project it was our intention to run all models until they had precessed by $180^\circ$. 
For models 3 and 4, however, we found that the discs become eccentric on a timescale of approximately 60 orbits (16 binary orbits), leading to undesirable changes in the disc structure due to our use of an open inner boundary condition.
Thus we did not evolve these models until they had precessed by $180^\circ$. We note that the timescale for the observed disc eccentricity growth corresponds closely to that obtained by \citep{kley-pap}. 
Model 5 did not suffer from this problem because the lower inclination angle induced more rapid precession, such that this model could be evolved until it had precessed by $180^\circ$, but prior to growth of significant eccentricity. 
The study of disc eccentricity goes beyond the scope of this paper, but the results for models 3 and 4 suggest that very long term integrations may lead to the formation of inclined and eccentric discs.
The fact that we did not evolve models 3 and 4 through $180^\circ$ of precession means that the upper and middle panels of Fig.\ref{images51} display images which correspond to $90^\circ$ of precession.

It is clear from these
panels that the disc in Model 3 has precessed as a rigid
body, with almost no twist apparent. Consequently, after
precessing through $90^{\circ}$ the disc appears edge on
when projected into the (${\bf e}_1$, ${\bf e}_3$) plane
and face on when projected into the (${\bf e}_2$, ${\bf e}_3$) plane.
The middle right panel
of Fig.~\ref{images51}, however, shows that the disc in Model 4
has developed a small twist due to the inner disc not having
precessed by $90^{\circ}$ at the same moment when the outer disc has.
The lower panels of Fig.~\ref{images51} show column density plots
for Model 5, which was evolved until the outer disc had precessed
by $180^{\circ}$ (assisted by the fact that the precession is
faster in this case). It is apparent from these panels that
the disc has developed a small twist since the inner disc has not
precessed a full $180^{\circ}$ by the end of the simulation. 

The precession angles for Model 3 are plotted in the top left
panel of Fig.~\ref{full51}. The blue solid line corresponds to
disc material orbiting between radii $r=$ 8 -- 9,
the red dashed line radii between $r=$ 1 -- 2,
and the black dotted line to radii between $r=$ 4 -- 5.
A black solid line is also plotted showing the precession
angle integrated over the whole disc. The fact that these lines
effectively lie on top of each other shows that the disc
precesses as a rigid body with essentially no twist. Rigid-body
precession is expected in this case since the warp propagation time
is less than the differential precession time (see Table 1).

\begin{figure*}
\begin{center}
\resizebox{10cm}{8cm}{\includegraphics[width=100mm,angle=0]{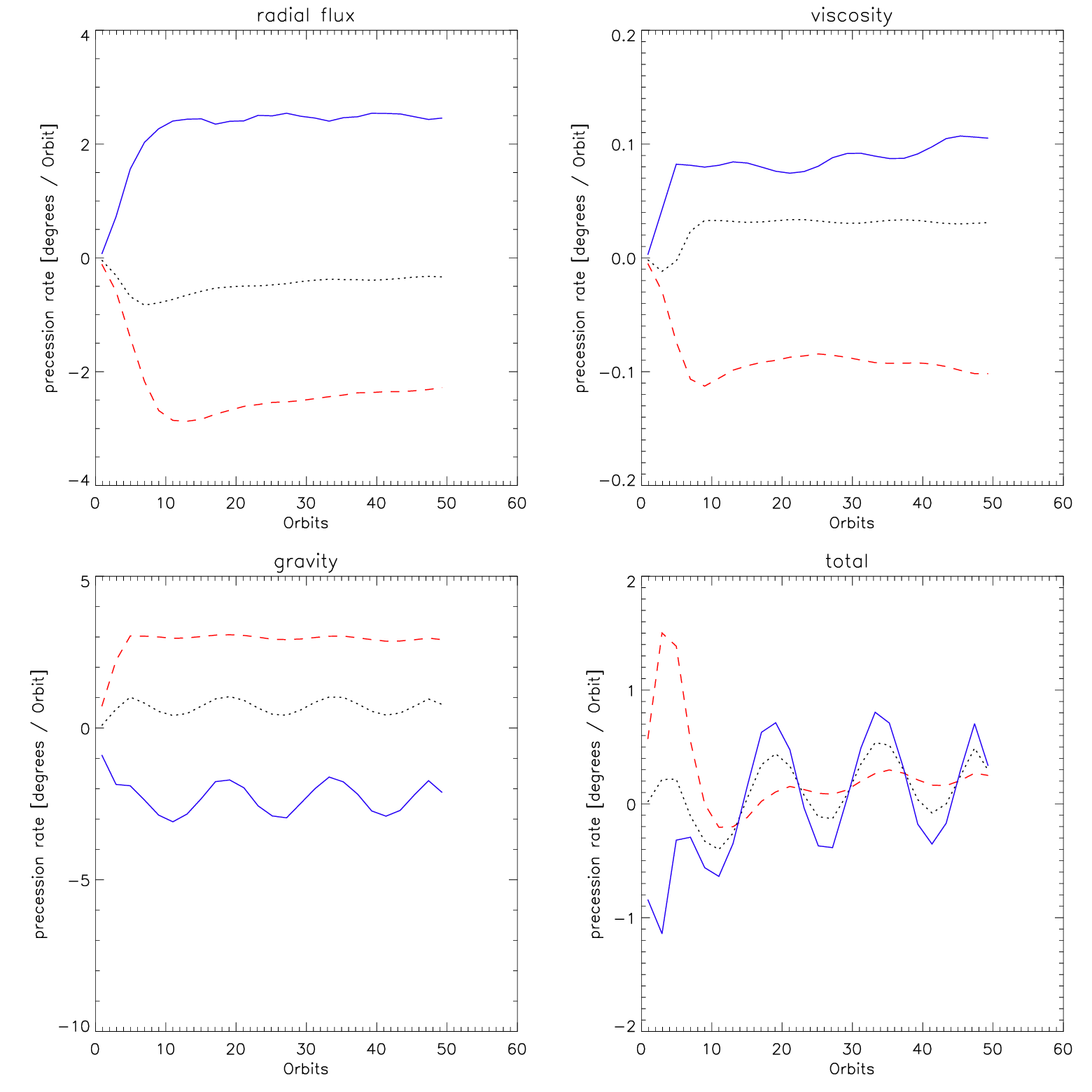}}
\caption{Precession rates due to the contributions discussed in Sect. 3.6
for Model 5. The blue solid line corresponds to disc material between 
$r=8$ and $r=9$, the red dashed line to disc material between 
$r=1$ and $r=2$ and the black dotted line to disc material between 
$r=4$ and $r=5$. These rates have been calculated using Eqs.~(\ref{rates}). In the lower right panel the total rate is displayed, which has beeen calculated using Eq.~(\ref{totrate}), where we subtracted the constant mean precession rate $\Omega_F$ of the precessing frame.}
\label{xtorque51}
\end{center}
\end{figure*}

\begin{figure*}
\begin{center}
\resizebox{10cm}{8cm}{\includegraphics[width=100mm,angle=0]{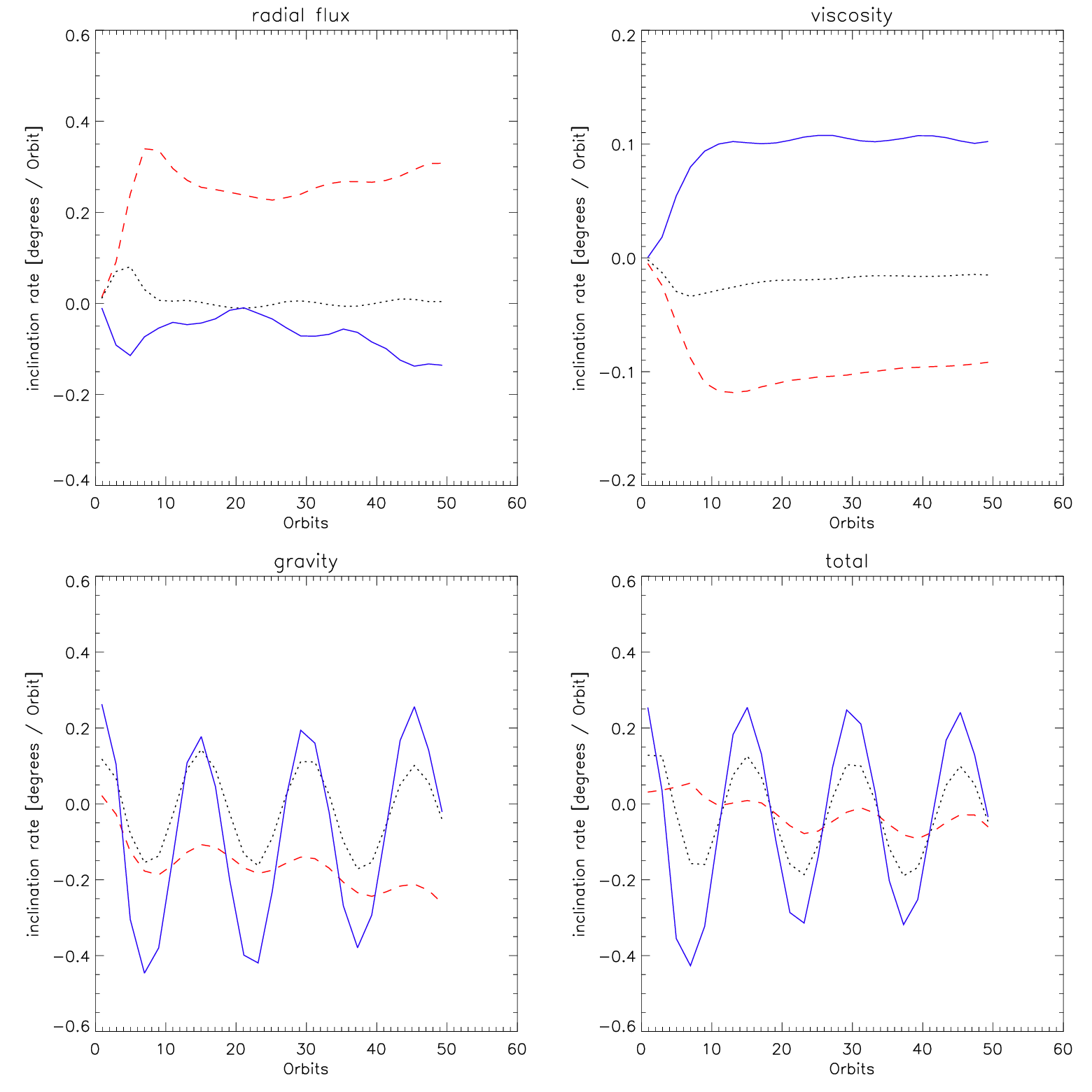}}
\end{center}
\caption{Inclination rates due to the contributions discussed in
Sect. 3.6 for Model 5. The blue solid line corresponds to disc material between 
$r=8$ and $r=9$, the red dashed line to disc material between 
$r=1$ and $r=2$ and the black dotted line to disc material between 
$r=4$ and $r=5$. These rates have been calculated using Eqs.~(\ref{rates}).}
\label{ytorque51}
\end{figure*}

The inclination angles for Model 3 are plotted in the top right
panel of Fig.~\ref{full51}. The line styles and colours correspond
to the same disc annuli as described above. It is clear that the
inclination evolution in this case is very slow. The global viscous
evolution time for this disc is approximately 10,000 orbits, and 
a linear extrapolation of the inclination evolution shown
in Fig.~\ref{full51} gives a time of $\sim 13500$ orbits for
the disc to fully align with the binary orbit. Clearly the
disc inclination is evolving on the viscous time in this case.

The evolution of the precession angles for Model 4 are shown in
the left middle panel of Fig~\ref{full51}.
The line styles and colours correspond to disc annuli as
described above. We see that this disc develops a well-defined
twist which is set up within the first 5 -- 10 orbits due to
the inner and outer parts of the disc undergoing differential
precession, with the outer disc precessing faster than the inner disc.
The magnitude of the twist is approximately $12^{\circ}$.
Thereafter the disc is seen to precess as a rigid body,
maintaining the same twisted shape. We note that this disc
is expected to achieve rigid body precession from
consideration of the warp propagation and differential
precession times (see Table 1).

The inclination angles for Model 4 are shown in the right middle
panel of Fig.~\ref{full51}. We see that the disc in this case has
developed a small warp, with the inner and outer discs having a
difference of inclination of about $1^{\circ}$. We also observe
that the inclination is evolving more rapidly than is the case
for Model 3. The global viscous evolution time for this disc
is approximately 1500 orbits, and extrapolating the inclination
evolution observed in Fig.~\ref{full51} gives an estimated
time for alignment of $\sim 1000$ orbits, close to the expected value. 

It is interesting at this point to compare our results for
Models 3 and 4 with Model 9 presented by \cite{larwood},
which had comparable parameters ($h=0.03$, $D/R=3$, 45 degree inclination).
Although it is difficult to know the precise value of the $\alpha$ viscosity
operating in the SPH simulations, it is likely that the effective
$\alpha$ value lies between the values used in our Model 3 ($\alpha=0.015$)
and Model 4 ($\alpha=0.1$). Nonetheless, we find qualitatively different
behaviour, with our models quickly achieving a state of rigid body
precession, with a small twist and warp which vary smoothly
across the disc. Model 9 of \cite{larwood}
shows significant disruption of the disc, which effective breaks 
discontinuously into two distincts parts which become mutually
inclined by approximately $25^{\circ}$. The origin of this discrepancy
is unclear, but one possibility is that SPH simulations
using 20,000 particles are only marginally able to resolve the
vertical structure of a disc with $h=0.03$. This may have the effect
of reducing the efficacy of warp propagation below that which
is expected from consideration of the disc physical parameters.
\begin{figure}
\begin{center}
\resizebox{7cm}{6cm}{\includegraphics[width=100mm,angle=0]{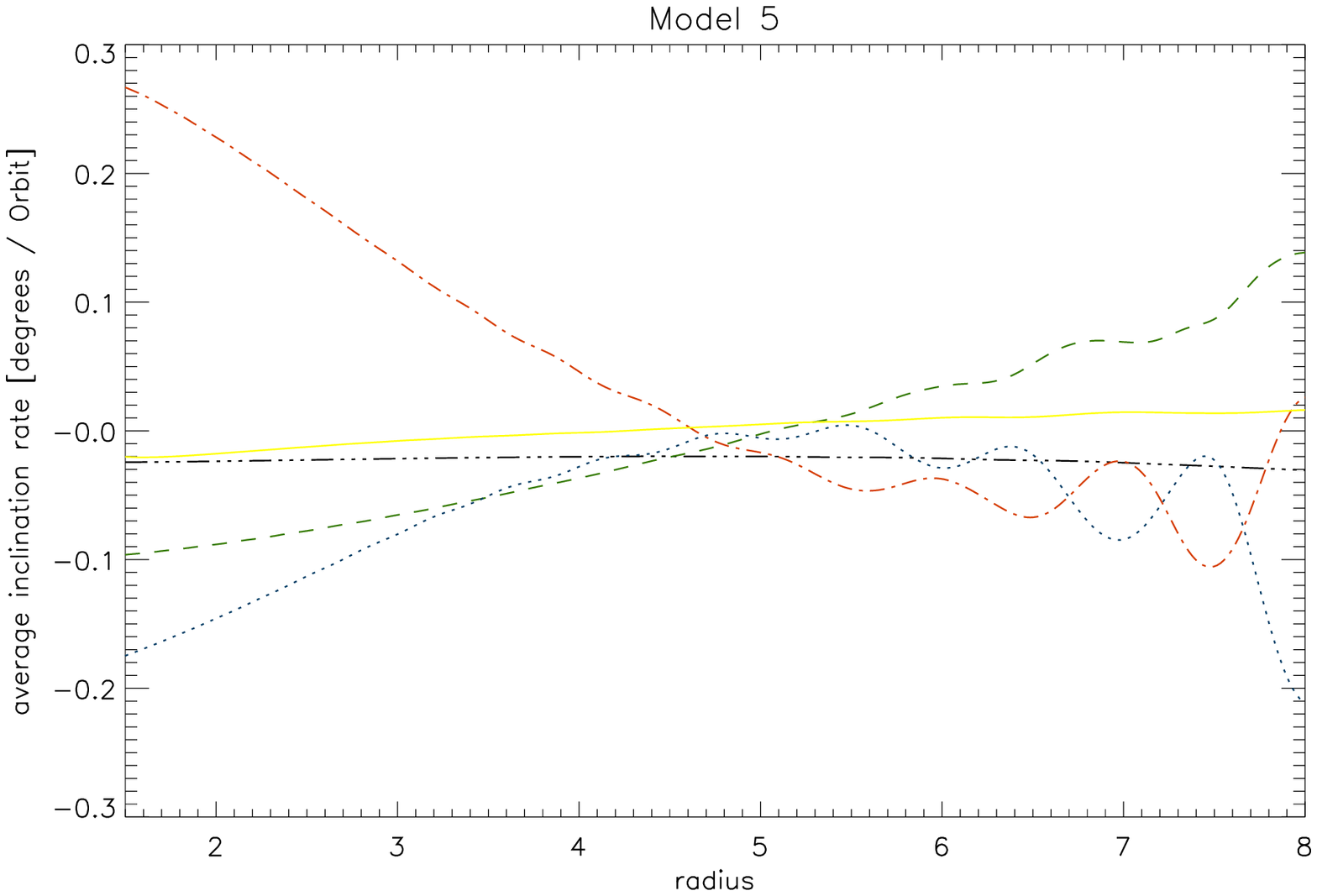}}
\end{center}
\caption{Time averaged inclination change rates as
function of radius for Model 5.
Red dashed-dotted line: radial flux of angular momentum;
blue dotted line: gravitational interaction with the companion star;
green dashed line: viscous friction between adjacent radial disc shells;
black dashed-triple-dotted line: total inclination change rate.}
\label{xaverage_A}
\end{figure}

%% projectional plot model5
%% prec2 plot model 5
We now turn to Model 5. The precession angles for this run are displayed
in the lower left panel of Fig.\ref{full51}.
Given that the precession frequency,
given by Eq.~(\ref{omega}), is proportional to $\cos{(\gamma_F)}$,
the disc precesses faster in this simulation than in Models 1 -- 4,
and this faster precession can be observed in Fig.~\ref{full51}.
The lower left panel of this figure shows the precession angles as
a function of time for disc annuli located between $r=$ 1 -- 2,
$r=$ 4 -- 5, and $r=$ 8 -- 9, with the linestyles and colours being the
same as described above. 
The larger precession frequency induces greater differential precession
in the disc, and the consequence of this is that Model 5 shows a
slightly larger twist ($15^{\circ}$) than Model 4 ($12^{\circ}$).
Also plotted in this panel are precession
angles that would be observed if the inner (black dashed-dotted line)
and outer (black dashed-triple-dotted line) disc annuli precessed
at their free particle rates given by \citep{papterquem}:
\begin{eqnarray}
\omega_Z=-\frac{3}{8\Omega_K}\frac{GM_S}{D^3}(3\cos^2{\delta}-1).
\label{free}
\end{eqnarray}
As can be seen from the figure, the disc parts tend initially
towards their free particle rates, such that the outer disc
precesses faster than the inner disc. Information about this
differential precession is communicated across the disc, and internal
stresses are established which cause the inner disc precession
to speed up and the outer disc precession to slow down, such that
precession at a uniform rate occurs after approximately 10 orbits.
The warp communication and differential precession
times given in Table 1 for this model predict rigid-body precession,
as observed.
%% xtorque51c plot
The emergence of internal stresses as the disc twists up
is illustrated in Fig.\ref{xtorque51}, where the different
physical contributions to the precession rate are displayed for Model 5.
These have been calculated according to the procedure outlined in
Sect. 3.6. Here the different linestyles correspond to the same disc
annuli described above (solid blue line: $r=$ 8 -- 9; dotted black line:
$r=$ 4 -- 5; dashed red line: $r=$ 1 -- 2).
From the lower left panel of Fig.\ref{xtorque51}, we can 
observe that the gravitational interaction with the secondary
causes a negative precession rate for the outer annulus, and a
positive precession rate for the inner annulus, and these
rates correspond to the free particle precession rates.
Since the frame in which these rates were measured
precesses in a retrograde sense with a mean frequency $\Omega_F<0$,
this means that the inner annulus would lag behind while the outer
annulus would lead the precession of the reference frame, 
if the companion's gravity was the only contributor to the 
total precession rate. In fact, this is what is observed until
about $t=10$ orbits, during which time the disc precesses differentially
(see Fig.\ref{full51}, lower left panel).
%% explaining rigid behaviour
As the disc starts to develop a twist due to differential
precession, misalignment of disc midplanes as a function of radius
causes pressure forces to generate horizontal shear motions that
drive the propagation of bending disturbances, and these
communicate information about precession angle changes across the disc.
The resulting internal stresses are clearly proportional
to the degree of twisting, allowing the disc to achieve a 
state of rigid body precession once a sufficiently twisted disc
has been established. At this point the disc structure becomes
quasi-static in the precessing frame as each disc annulus
precesses at the same rate.
This sequence of events is demonstrated by the upper left
panel of Fig.\ref{xtorque51}, which shows the contribution to
the precession rate from the pressure-induced radial flux
of angular momentum. After $t=10$ orbits, 
the contribution to the precession rate from the radial
flux almost completely compensates the effect of gravity,
such that the total precession rate (relative to the
precession of the reference frame) is zero for all disc annuli,
as can be seen in the lower right panel of Fig.\ref{xtorque51},
which shows the sum of the contributions to the precession rate.
This demonstrates that the disc manages to redistribute
the angular momentum such that rigid body precession is achieved.
Once again we note that in this figure we have subtracted the 
mean precession rate $\Omega_F$ of the precessing frame.
%% effect of viscosity

The effect of viscosity as calculated in Sect. 3.6 is
depicted in the upper right panel of Fig.\ref{xtorque51}. We observe that
the effect of viscous friction between differentially precessing radially
adjacent disc shells causes the disc to precess more rigidly.
However, the dominant effect of
the viscosity is the damping of the horizontal shear motions induced
by the twist.
Since these shear motions are responsible for driving the bending
disturbances, the viscosity will lead to weakened communication, and
the redistribution of angular momentum due to radial advective
fluxes shown in the upper left panel of Fig.\ref{xtorque51} will be suppressed.
This effect is dominant and opposite to the one seen in the upper right panel of
Fig.\ref{xtorque51}.
%% ytorque51

The inclination angles for Model 5 are shown in the lower right 
panel of Fig.~\ref{full51}. It is clear that the disc develops a
small warp, with the inner disc having $\sim 1^{\circ}$
higher inclination than the outer disc. The rate of global inclination change
is similar to that observed in Model 4, suggesting alignment of
the disc in the global viscous evolution timescale.

We plot the different physical contributions to the rate of
inclination change for Model 5 in Fig.\ref{ytorque51}.
In Fig.\ref{xaverage_A} we also display the time averaged contributions to
the rate of inclination change as a function of radius, where the
time averaging was performed over the full length of the simulation.
As may be observed in the lower left panel of Fig.\ref{ytorque51},
and even more clearly in Fig.\ref{xaverage_A} (blue dotted line),
the companion's gravity induces a globally negative net inclination
change, leading to coplanarity of the entire disc on the viscous time scale,
in agreement with the lower right panel of Fig.\ref{full51}.
This is consistent with the findings of
\cite{papterquem} that the zero-frequency (secular)
gravitational interaction will lead to coplanarity. 
We note that because of the time averaging, this is the dominant
gravitational contribution that we pick up in
Fig.\ref{xaverage_A} (blue dotted line). 

More interestingly, we can observe from the upper
left panel of Fig.\ref{ytorque51}, and in 
Fig.\ref{xaverage_A} (red dashed-dotted line), 
that the contribution from the radial fluxes not only counteracts the 
effect of gravity, it actually overshoots slightly, 
causing the inner disc to become more inclined than the outer disc.
This change of inclination is seen in the upper right panel of
Fig.\ref{full51}. Thus the communication of bending disturbances
on the one hand leads to rigid precession of the disc,
and on the other hand causes the disc to become slightly warped.
Viscosity tends to reduce the amount of warping, 
such that it acts to flatten the differential inclination profile
across the disc radius, 
as can be seen from the green dashed line in Fig.\ref{xaverage_A}.
Thus the total average inclination rate in Fig.\ref{xaverage_A} 
(dashed-triple-dotted line) is negative for the entire disc,
while the magnitude is slightly bigger for the outer disc,
with the consequence that the disc has developed a slight warp and
tends to become coplanar with the orbital plane on the viscous timescale,
as seen in the lower right panel of Fig.\ref{full51}.
%% explaining model 6 - prec2 for model 6

\subsection{Models 6 and 7}
\begin{figure*}
\begin{center}
\resizebox{13.33cm}{23cm}{\includegraphics[width=100mm,angle=90]{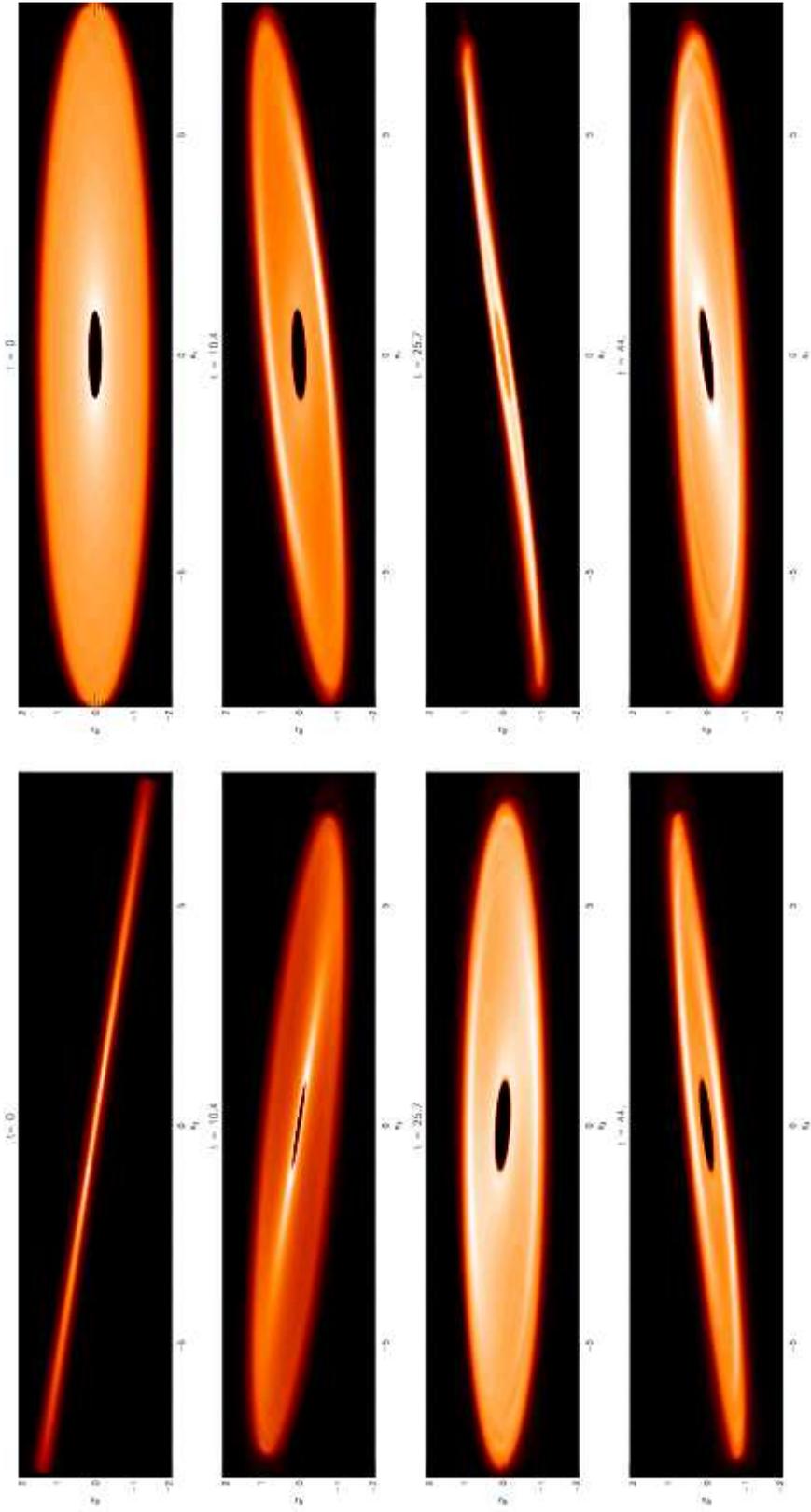}}
\end{center}
\caption{Column density plots for Model 6. The left-hand panels
are projections onto the (${\bf e_2}$, ${\bf e_3}$) plane,
and the right-hand panels are projections onto the
(${\bf e_1}$, ${\bf e_3}$) plane. The disc structure is displayed at
the different times indicated.}
\label{image50a}
\end{figure*}
\begin{figure*}
\begin{center}
\resizebox{18cm}{10.5cm}{\includegraphics[width=100mm,angle=0]{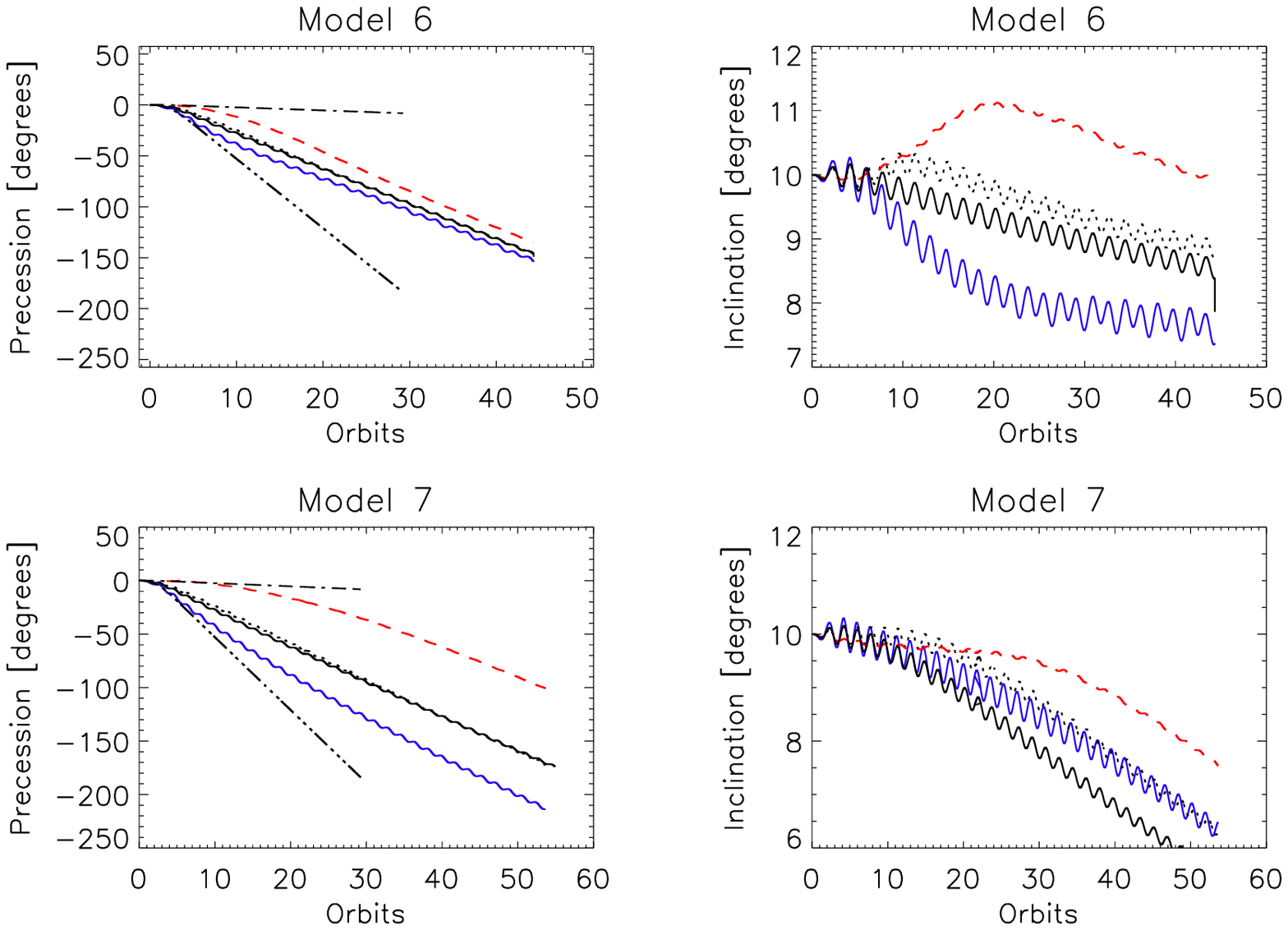}}
\end{center}
\caption{Evolution of the precession (left panels) and inclination angles
(right panels). The blue solid line corresponds to disc material between
$r=7$ and $r=8$, the red dashed line to disc material between
$r=1$ and $r=2$, and the black dotted line to disc material
between $r=4$ and $r=5$. The solid black line represents the entire disc.}
\label{full50}
\end{figure*}
We now discuss Models 6 and 7, for which the discs have aspect ratio $h=0.01$,
an inclination angle of $10^{\circ}$, and were performed in a
non precessing frame. Model 6 has $\alpha=0.005$,
so warps propagate {\it via} bending waves in this case, and Model 7 had
$\alpha=0.1$, so warps propagate diffusively in this model.
We see from Table 1 that differential precession 
is expected to disrupt the discs in Model 7,
as the warp propagation time is
longer than the differential precession time. Rigid-body
precession is expected for Model 6.

Column density plots showing the disc in Model 6 are presented
in Fig.\ref{image50a} for different times during the evolution. 
The top panels show the initial state of the disc. 
The disc appears edge on in the (${\bf e_2}$, ${\bf e_3}$) plane and face 
on in the  (${\bf e_1}$, ${\bf e_3}$) plane. As we move down to the 
next panels the outer edge of the disc has precessed by 45 degrees at 
time $t=10.4$, as can be seen in Fig.\ref{full50} (upper left panel). 
The inner region of the disc, however, has only precessed by about 12 degrees 
at this stage and so appears almost edge-on in the left panel.
In the third panels down the outer disc has precessed by 
90 degrees at $t=25.7$ and appears edge on in the 
(${\bf e_1}$, ${\bf e_3}$) plane and face on in the 
(${\bf e_2}$, ${\bf e_3}$) plane. The inner disc has precessed by
approximately 70 degrees by this time, and so does not appear edge-on
in either projections.
The bottom panels show the disc structure at the final output of the 
simulation which was halted at $t=44$ orbits, by which time the
outer disc had precessed by approximately $160^{\circ}$.

The precession angles for Model 6 are shown in the upper left panel
of Fig.~\ref{full50}. The blue solid line corresponds to disc material
orbiting in an annulus between $r=$ 7 -- 8, the black dotted line
corresponds to the annulus between $r=$ 4 -- 5, and the red dashed
line corresponds to the annulus between $r=$ 1 -- 2.
As with Models 4 and 5, we see that the disc inner and outer 
annuli begin to precess at the local free-particle rates,
and the disc develops a significant twist which
reaches a maximum amplitude of approximately $30^{\circ}$.
As the disc becomes increasingly twisted, however, internal
stresses are established which cause the inner disc
precession rate to increase and the outer disc precession rate
to decrease. Eventually the disc reaches a state of rigid body
precession after a time of about 10 orbits. After this time
we see that there is a long term readjustment of the degree
of twist such that by $t=44$ orbits the twist angle has reduced 
from $\sim 30^{\circ}$ to $\sim 20^{\circ}$.

\begin{figure}
\begin{center}
\resizebox{7cm}{6cm}{\includegraphics[width=100mm,angle=0]{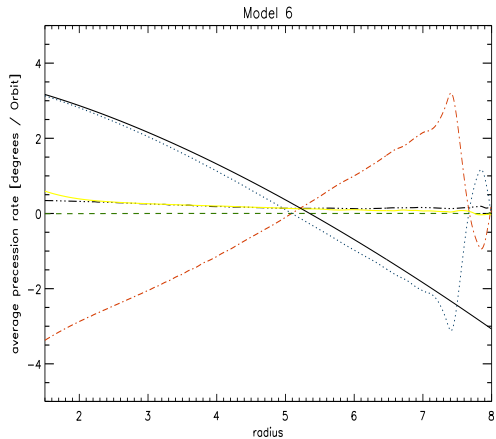}}
\end{center}
\caption{Average precession rates as function of radius for Model 6.
Red dashed-dotted line: radial flux of angular momentum;
blue dotted line: gravitational interaction with the companion star;
green dashed line: viscous friction between adjacent radial disc shells;
black dashed-triple-dotted line: total inclination change rate.
The black solid line represents the precession rate of free particles.}
\label{xaverage_C}
\end{figure}

In Fig.~\ref{xaverage_C}
we display the time integrated contributions to the precession rate
as a function of radius for Model 6, following the procedure described
in Sect. 3.6. Note that the average (negative) precession rate of the disc
has been subtracted in this figure, such that a positive value
corresponds to precession that will lag that of the overall disc, and
a negative value corresponds to precession that leads. 
The contribution from the companion's gravity
is shown by the blue dotted line, and this is very similar
to the free-particle precession rate which is plotted using the
black solid line. The red dashed-dotted line shows the contribution
from the pressure-induced radial flux of angular momentum, and 
this is seen to almost exactly counterbalance the effect of gravity,
showing that it is the pressure-induced radial motion in the disc
which is responsible for halting the differential precession
and inducing rigid body precession.
The contribution from the disc viscosity is shown by
the green dashed line, and is seen to be negligible.

\begin{figure}
\begin{center}
\resizebox{7cm}{6cm}{\includegraphics[width=100mm,angle=0]{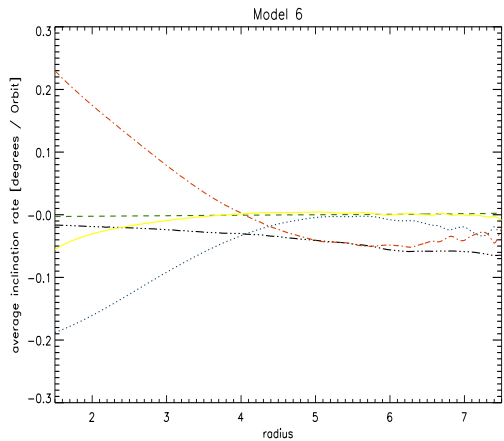}}
\end{center}
\caption{Average inclination change rates as function of radius
for Model 6. Red-dashed-dotted line: radial flux of angular momentum;
blue dotted line: gravitational interaction with the companion star;
green dashed line: viscous friction between adjacent radial disc shells
black dashed-triple-dotted line: total inclination change rate.}
\label{xaverage_B}
\end{figure}

The inclination angles for Model 6 are shown in the top right panel
of Fig.~\ref{full50}, where the line styles and colours correspond
to the disc annuli described above. As with Model 3, we see that
the disc develops a warp such that the inner disc has a larger
inclination to the binary plane than the outer disc. Indeed, the
inclination of the inner disc is seen to increase initially,
before decreasing after approximately 20 orbits as the disc as
a whole aligns with the binary orbit plane. As discussed previously,
disc alignment is expected to occur on the viscous time scale,
which for Model 6 is approximately $3 \times 10^5$ orbits. It is obvious
from Fig.~\ref{full50}, however, that the disc in our simulation is
not aligning on such a long time scale, but will actually align
within $\sim 650$ orbits. Possible reasons for this enhanced alignment
rate are discussed below.

Fig.~\ref{xaverage_B} shows the time integrated contributions to
the rate of inclination evolution as a function of radius for Model 6.
We see that the companion's gravity causes the inclination to decrease
in the disc inner and outer regions, and in the inner regions the
radial flux of angular momentum opposes this and increases
the inclination during the simulation. Indeed the radial flux 
contribution causes the inclination angle to grow there, as observed
in the lower right panel of Fig.~\ref{full50}. The outer parts of
the disc experience negative contributions from both the companion's
gravity and the pressure-induced radial flux, driving alignment of
the disc on the relatively short timescale described above. It is
clear that the disc viscosity plays a negligible role in driving the
inclination evolution (green dashed line).

We may compare our model 6 with model c in Figure 7 of \citet{lubow},
which had $H/R \simeq 0.012$, $\alpha=0.01$ and $D/R=0.3$. 
Detailed comparison between the results presented in \citet{lubow}
and our model 6 is not possible, but we note that 
there is general agreement in the prediction that a disc with
$H/R \simeq 0.01$ and $\alpha \lesssim H/R$  will show significant
distortion due to twisting and warping. A detailed comparison
between our non linear simulation results and the predictions of
linear theory will be presented in a future publication.
\begin{figure*}
\begin{center}
\resizebox{13.33cm}{23cm}{\includegraphics[width=100mm,angle=90]{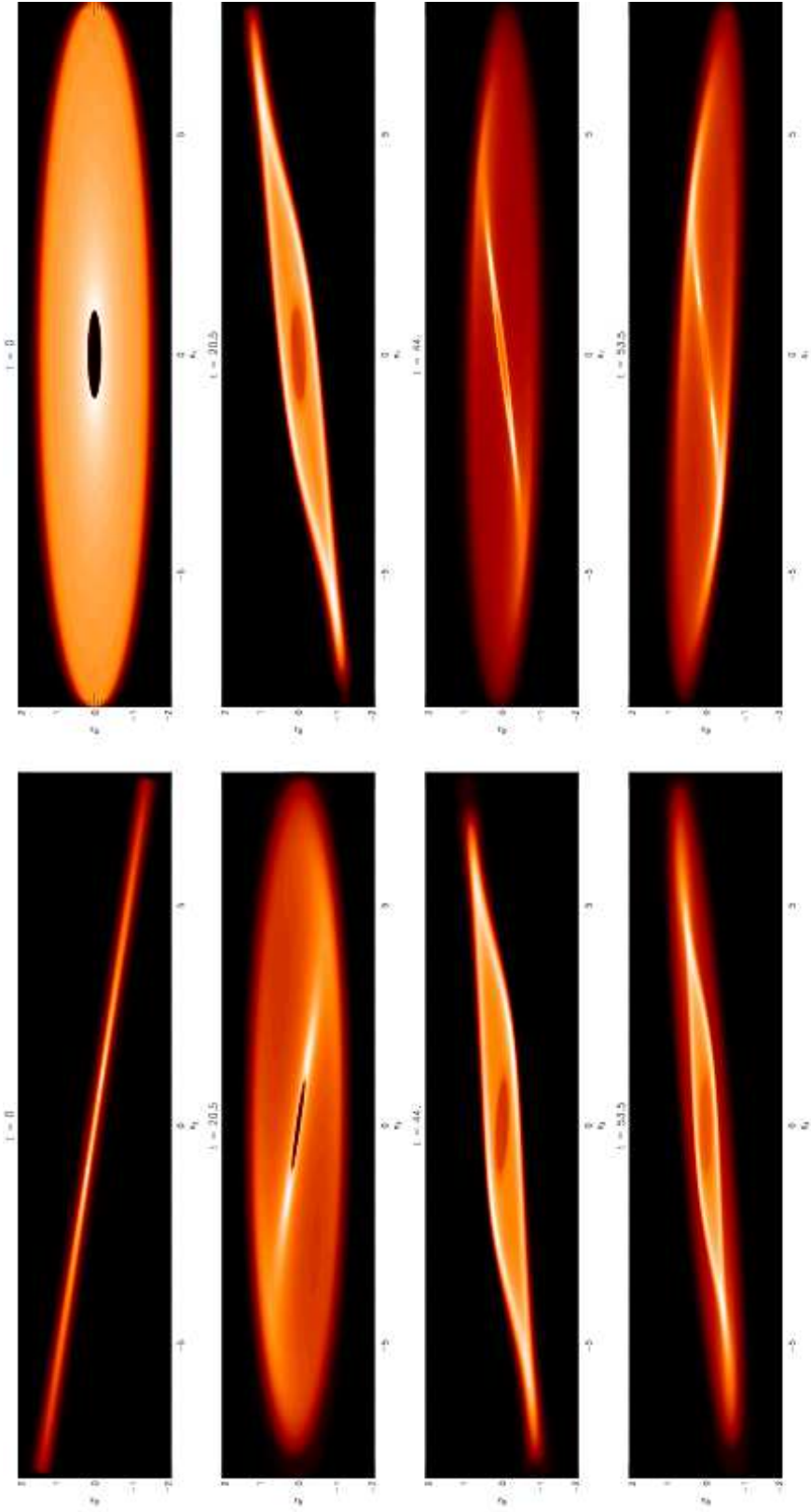}}
\end{center}
\caption{Column density plots for Model 7. The left-hand panels
are projections onto the (${\bf e_2}$, ${\bf e_3}$) plane,
and the right-hand panels are projections onto the
(${\bf e_1}$, ${\bf e_3}$) plane. The disc structure is displayed at
the different times indicated.}
\label{image50b}
\end{figure*}

We now discuss the results from Model 7.
In Fig.\ref{image50b} we display column density plots for this model.
The top panels show the initial state. The second panels down
show the disc at time $t=20.5$. At this stage the outer disc 
edge has precessed by 90 degrees and appears edge on in the 
(${\bf e_1}$, ${\bf e_3}$) plane, and face on in the (${\bf e_2}$, ${\bf e_3}$) 
plane. By contrast, the inner disc has not precessed very much 
and still appears almost edge on in the (${\bf e_2}$, ${\bf e_3}$) plane 
and face on in the (${\bf e_1}$, ${\bf e_3}$) plane. 
At time $t=44$ orbits the outer disc has precessed by 180 degrees and appears 
edge on in the (${\bf e_2}$, ${\bf e_3}$) plane and face on in 
the (${\bf e_1}$, ${\bf e_3}$) plane. 
The inner disc, however, has only precessed by 
about 70 degrees by this time, and so does not appear edge on in either
projection. The bottom panels show the final stage of the simulation
when the outer disc has precessed by about 220 degrees. It is clear
from these images that the disc in Model 7 has adopted a highly twisted shape,
but one in which the precession angle for different disc
annuli varies very smoothly across the disc. 

The precession angles for Model 7 are displayed in the
lower left panel of Fig.\ref{full50}. We observe that the 
disc is in a state of differential precession at the end of the simulation,
and the twist between inner and outer disc annuli is approximately 110 degrees 
at this stage. Although this twist is large, the actual rate of differential
precession at the end of the simulation is very small, as the
precession rates of the inner and outer disc have converged during the
run. Indeed, extrapolation of the precession angles in
Fig.~\ref{full50} indicates that rigid body precession will be 
achieved at a time equal to $t=56.6$ orbits, at which point
the twist amplitude will be $111.7^{\circ}$. This suggests that
even though the warp propagation time is much smaller than the
differential precession time for Model 7 (see Table 1), internal
stresses can be set up within the disc that cause uniform precession
once the disc becomes highly twisted. Adopting a definition of disc
disruption which requires the twist to become greater than 180 degrees,
we suggest that the disc in Model 7 will not be disrupted, but  
will instead adopt a highly twisted but uniformly precessing
configuration.

The inclination angles for Model 7 are shown in the lower right panel
of Fig.~\ref{full50}, where the lines styles correspond to
the disc annuli described above.
We can see that the disc in this case
develops a small warp with the difference between the
inclinations of the inner and outer disc being $\sim 1^{\circ}$.
We also note that it appears that the integrated tilt of the 
whole disc is actually smaller than for any of the individual
disc annuli, but this is an effect of averging over a 
significantly twisted disc. 
As observed for Model 6, we see that the inclination of the
disc evolves rather quickly. The global viscous evolution time
for Model 7 is $\sim 15000$ orbits, whereas extrapolation of the
inclination angles for either the inner or outer disc suggests
that they will completely align after $\sim 100$ orbits.

What is the explanation for this rapid alignment observed for Models 6 and 7 ?
One possibility that we have explored is that numerical diffusion may
cause a tilted disc to align with the equatorial plane of the 
computational grid,
since the advection routine does not conserve total angular momentum when
the disc azimuthal velocity is not directed along the azimuthal
direction of the computational grid. Lower resolution test calculations 
performed for a tilted disc without a companion indicate 
that this effect can cause eventual disc alignment, but at a rate which
is at most 7 times smaller than observed in Models 6 and 7.
This suggests that numerical diffusion is not the cause
of the rapid alignment.

\begin{figure}
\begin{center}
\resizebox{8.8cm}{5cm}{\includegraphics[width=100mm,angle=0]{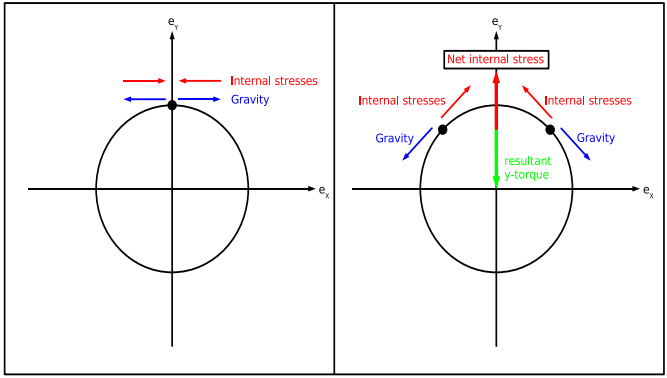}}
\end{center}
\caption{Diagram illustrating the alignment torques which
arise in highly twisted discs, as discussed in the text.}
\label{align}
\end{figure}

Another possibilty arises when we consider the evolution of a 
rigidly precessing, tilted disc which is in a precessing 
reference frame whose precession frequency is equal to that
of the disc.
Consider first the situation in which the angular momentum 
vectors for all disc annuli lie in the $y-z$ plane of the binary
frame initially (i.e. the $z$ direction of the precessing
frame lies initially in the $y-z$ plane of the binary frame,
pointing along the positive $y$ and $z$ directions). 
A torque exerted on disc annuli in the $y$ direction,
(a `$y$-torque), will cause the disc inclination to change.
A torque exerted in the $x$ direction (an `$x$-torque'),
will cause precession at a rate which differs from the 
precession rate of the frame. A positive $x$-torque causes precession
to occur at a rate which is faster than the frame, and a
negative $x$-torque causes slower precession. This
general picture is expressed mathematically by Eqs.~(\ref{rates}),
and is shown diagramatically in the left hand image of Fig.~\ref{align}.
For a disc which precesses uniformly without any twist, then
the companion's gravity will exert positive $x$-torques on the
outer disc annuli, and negative $x$-torques on the inner disc annuli,
as it tries to cause the disc to precess differentially.
Internal stresses, however, will exactly counterbalance these $x$-torques to maintain uniform precession.
In this simple scenario there is no torque exerted on the disc in the
$y$ direction to modify its inclination.

Consider now the case of a highly twisted disc which is being
forced to precess uniformly, and whose annuli all have 
the same inclination. Gravity will again try to induce
differential precession, but now the torque exerted on
disc annuli will have both an $x$ and a $y$ component
due to the twist. Resolving the vectors associated with
the internal stresses required to enforce uniform precession
shows that their combined $x$-torques operate in opposite
directions such that they cancel when integrated
over the disc. The combined $y$-torques, however, 
operate in the same direction, implying that the internal
torques generate a net $y$-torque on the disc. Conservation of
angular momentum obviously prohibits this from arising, 
suggesting that a $y$-torque must operate, whose effect is to
change the disc inclination. This is shown diagramatically
in the right panel of Fig.~\ref{align}.
Given that the internal 
torques are operating on the precession timescale
in order to maintain uniform precession, this argument suggests
that a highly twisted disc may align on the precession
time, as is observed for Models 6 and 7. Interestingly, the 
resultant $y$-torque that arises in this picture
is proportional to $\sin{(\Delta\beta)}$, where $\Delta\beta$ is
the twist amplitude, and the rate of inclination evolution
observed in Models 6 and 7 is in agreement with this scaling.

If the above argument is correct, then it suggests that
discs with modest levels of twist will align on the
global viscous timescale of the disc, but highly twisted discs
may align on the much shorter precession timescale.
This obviously has significant implications for astrophysical
systems which may be expected to harbour highly twisted discs,
such as the X-ray binaries Her X-1 and SS433.

\begin{figure*}
\begin{center}
\resizebox{10cm}{8cm}{\includegraphics[width=100mm,angle=0]{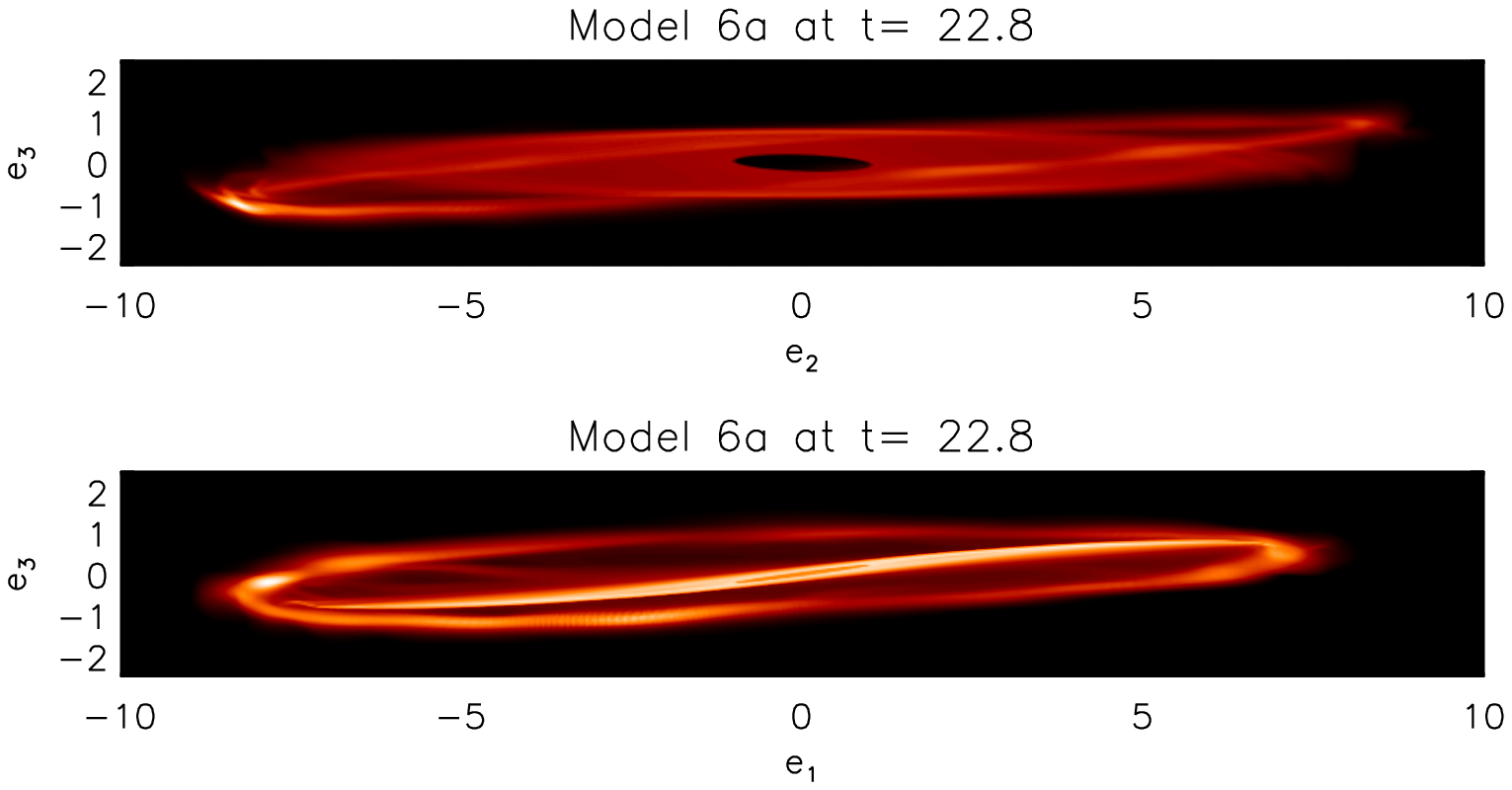}}
\resizebox{10cm}{8cm}{\includegraphics[width=100mm,angle=0]{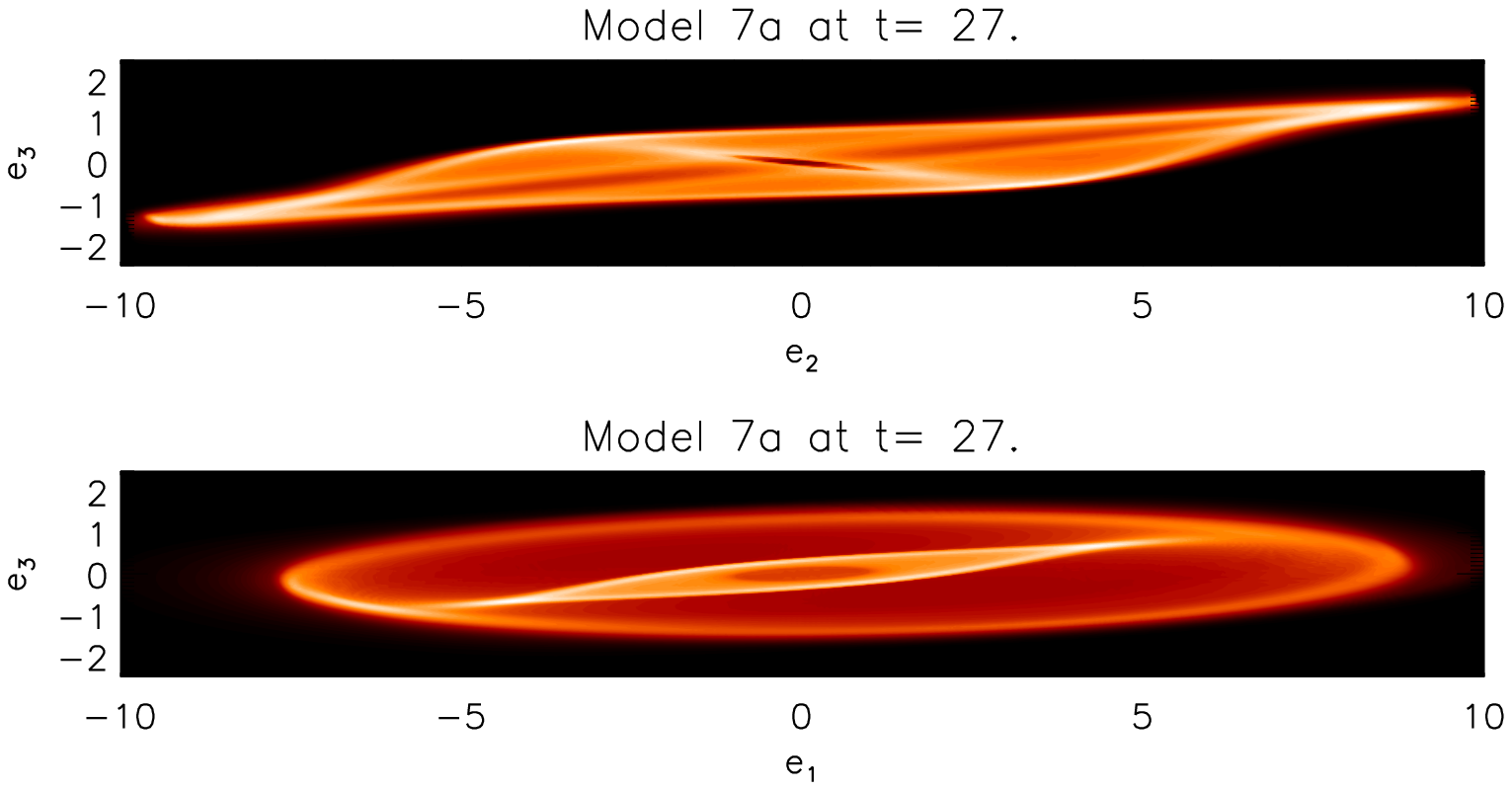}}
\end{center}
\caption{Column density plots of the two lower resolution runs Models 6a and 7a.
The upper panels show two different projections for Model 6a at time $t=22.8$
The lower panels show two different projections for Model 7a at time $t=27.0$.}
\label{image42}
\end{figure*}

\subsubsection{Models 6a and 7a - broken or disrupted discs ?}
We now briefly discuss the results of two lower resolution models
which were run during an early stage of this project: Models 6a and 7a.
These were identical to Models 6 and 7 except that the number
of grid cells in the radial direction was $N_r=300$, and the
disc outer radii were located at $R=10$ instead of $R=8$.
Extending the disc radius outward has the potential effect
of increasing the rate of differential precession across the
disc, provided that tidal truncation does not cause
the disc to shrink back down to $R=8$.

\begin{figure*}
\begin{center}
\resizebox{10cm}{8cm}{\includegraphics[width=100mm,angle=0]{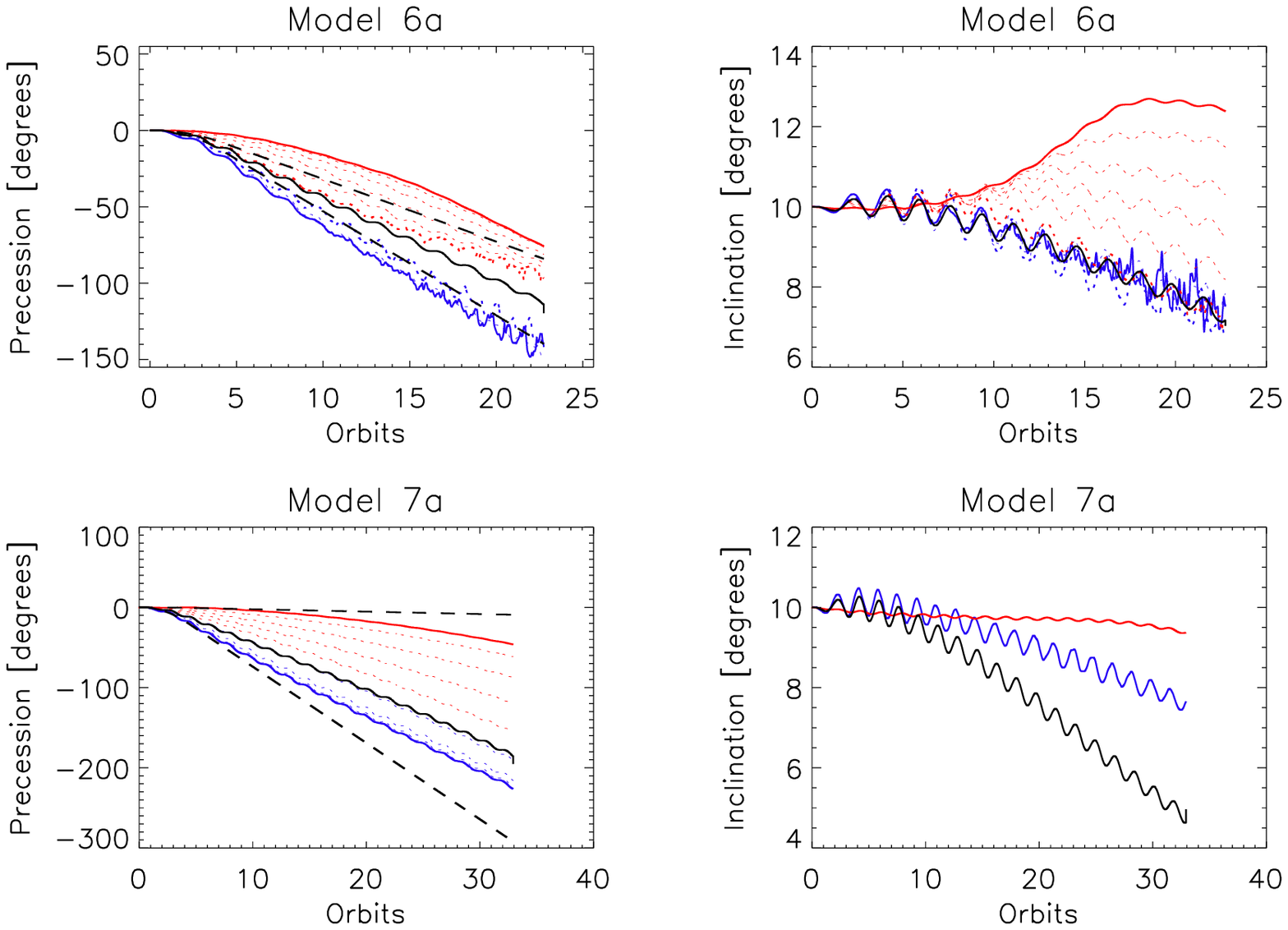}}
\end{center}
\caption{Evolution of precession (left panels) and inclination angles
(right panels) for Models 6a and 7a. The blue solid lines correspond
to disc annuli of unit radius lying between $r=7$ and $r=10$.
The solid blue line denotes the annulus lying between $r=9$ and $r=10$.
The red lines correspond to disc annuli of unit width
lying between $r=1$ and $=7$. The solid red line denotes the
annulus lying between $r=1$ and $r=2$.}
\label{full42}
\end{figure*}

Column density plots for Model 6a are shown in the top two panels
of Fig.~\ref{image42}, corresponding to an evolution time of $t=22.8$ orbits.
We can see that there has been some tidal truncation of
this disc, and the outer disc edge is significantly distorted
in the images. This arises because a narrow outer rim,
running between radii $r \sim 7$ and $r \sim 10$, 
has effectively detached itself from the main body of
the disc, and has evolved to have significantly different precession
and inclination angles compared to the main body of the disc.

The inclination and precession angles for Model 6a are displayed
in the upper left and right panels of Fig.~\ref{full42}, respectively.
The precession angle of the annulus which lies between
radii $r=$ 1 -- 2 is shown by the thick red solid line, and the
remaining red dotted lines show the precession angles for
disc annuli of unit width out to $r=7$.
The solid blue line corresponds to the disc annulus lying
between $r=$ 9 -- 10, and there are two additional blue dotted
lines for annuli lying between $r=$ 7 -- 8 and $r=$ 8 -- 9.
It is clear that the disc has separated into two parts discontinously
at a radius between $r=7$ and $r=8$, with the detached outer rim having
precessed significantly faster than the inner disc.
The inclination angles plotted in the upper right panel show
that the disc becomes significantly warped, with the outer rim tending
to align with the binary and the inner disc actually increasing
its inclination. Although the change in inclination appears to
occur fairly smoothly across the disc, it should be noted that
the disc has developed a fairly large warp of $5^{\circ}$.

The behaviour of Model 6a is reminiscent of the broken disc 
presented by \cite{larwood} (their broken Model 9 had $h=0.03$, 
inclination of $45^\circ$, $D/R=3$) except that the discontinuous break
in Model 6a occurs near to the edge of the disc. It would appear
that Model 6a breaks because the larger disc radius induces
a larger rate of differential precession, and also possibly because
warp propagation is retarded in the outer disc regions which are
subject to non linear density structures and shocks because
of strong interaction with the binary companion.

Column density plots for Model 7a are shown in the lower two panels
of Fig.~\ref{image42}, corresponding to a run time of $t=27$ orbits.
Here we can see that the disc has become very highly
twisted (the twist amplitude in the figure is $\sim 150^{\circ}$),
but does not show any of the discontinuous behaviour
observed in Model 6a. Instead the twist varies very smoothly across
the disc. The precession and inclination angles for Model 7a are
shown in the lower left and right panels of Fig.~\ref{full42}, 
respectively. The solid red line shows the precession angle of
the annulus located between $r=$ 1 -- 2, and the solid blue line
corresponds to the annulus between $r=$ 9 -- 10. By the end of the
simulation the twist amplitude is $\sim 170^{\circ}$,
and the precession rates of inner and outer disc remain very different
such that this disc will differentially precess through $180^{\circ}$.
So, by our working definition of disc disruption, 
this disc will be disrupted because of differential precession.
The inclination angles show that the disc also becomes significantly warped,
with the outer disc aligning with the binary orbit more rapidly
than the inner disc.

To summarise: Model 6a and 7a both show examples of discs which 
break or are disrupted because of differential precession.
In the case where warps propagate {\it via} bending waves,
the disc breaks discontinuously into two independently precessing
parts, but which themselves remain as coherent structures which
precess as rigid bodies.
In the case where warps propagate diffusively, viscosity is able
to smooth out any discontinuous behaviour, and instead the disc becomes
disrupted because the twist exceeds 180 degrees. In all probability
the disc annuli in Model 7a will continue to precess differentially
in the longer term, substantially destroying any coherent disc-like 
structure. The larger disc radius in this case, when compared
with Model 7, causes the differential precession rate across
the disc to be larger, and thus explains why Model 7a shows
disruption, whereas Model 7 results in a highly twisted disc which
is able to attain a state of rigid-body precession.

\section{Conclusions}
We have presented non linear simulations of accretion discs in close
binary systems in which the binary midplane is misaligned 
from the binary orbital plane. Previous work on this problem,
which employed linear theory, semi-analytic techniques and SPH simulations
\citep{papterquem, larwood, lubow}, suggests that under suitable conditions
the disc should become mildly warped and precess as a rigid-body
around the angular momentum vector of the binary system. The main aim
of the present study is to examine in detail the structure and evolution
of discs in misaligned binary systems, subject to variation of
the important physical parameters. We used the grid-based
code NIRVANA to perform the numerical simulations, and
in principle this should have the advantage over previous SPH simulation
studies that the disc viscosity is a well defined and controllable
parameter.

Following \cite{nelsonpap}, the propagation of linear bending 
waves was used as a means
to calibrate the code and test its ability to propagate
warps. Direct comparison with linear calculations show
that the code can propagate bending waves with a high
degree of accuracy for a range of disc models and warp
perturbation amplitudes.

In our study of discs in misaligned binary systems, a number of
different models were considered. Discs with aspect ratios 
$h=0.05$, 0.03 and 0.01 were studied, and for each of these
values models were computed with 
viscosity parameters $\alpha=h/2$ and $\alpha=0.1$.
The smaller value of $\alpha$ allows bending waves to propagate,
whereas the larger value causes warps to evolve diffusively.

It is expected that discs for which the warp propagation time, $\tau_W$,
is shorter than the differential precession time, $\tau_P$, will
undergo rigid-body precession. Discs which do not satisfy
this criterion may be disrupted by strong differential 
precession. All our models with $h=0.05$ and 0.03
have $\tau_W < \tau_P$, and the simulations show that each
of these discs attains a state of rigid-body precession.
Discs with lower viscosity maintained a precessing structure
with almost no discernible twist and warp,
due to the highly efficient warp propagation
associated with bending waves. Discs with larger viscosity
propagate warps less efficiently, and these models
underwent a short period of differential precession, setting
up a twist in the disc, prior to attaining a state of
rigid-body precession. 
Our analysis shows that internal
stresses are established in the disc as it becomes twisted,
which transport angular momentum across its radius and cause it
to precess rigidly. Models in which warp propagation is the
least efficient develop the largest twist, since the internal
stresses which hold the disc together against differential
precession are proportional to the twist amplitude.
Examination of the simulations shows that the dominant
contributor to these internal stresses are pressure-induced
radial angular momentum fluxes which are driven by local
misalignment of disc midplanes. 
When $\alpha=0.1$, the final amplitude of the
twist was just a few degrees for the $h=0.05$ disc, and 
for the $h=0.03$ disc it was between 12 -- 15 degrees,
depending on the binary inclination. The low viscosity
models showed essentially no twist.

In addition to becoming twisted, a disc may also become warped.
We find that for all models with $h=0.05$ and $h=0.03$
the degree of warping is very modest, with differences in
inclination across all disc annuli being less than 1 degree.
It is expected that these discs will also align with
the binary orbital plane on the global viscous evolution
time of the disc \citep{papterquem,larwood}, and all simulations with 
$h=0.05$ and $0.03$ are fully consistent with this expectation.

We also considered a number of thin disc models 
with $h=0.01$, where we not only varied the viscosity
but also the outer radius (taking values of
either $R=8$ or $R=10$). The low viscosity model 
with $R=8$ developed a significant twist of
between 20 -- $30^\circ$ before achieving a state
of solid body precession, as expected since
$\tau_W < \tau_P$ for this model.
A model run with $\alpha=0.1$ and $R=8$ was predicted to be
disrupted due to differential precession since $\tau_W > \tau_P$
in this case. During the simulation it developed a very large twist
$\sim 110^\circ$, but as the disc become increasingly twisted
the rate of differential precession decreased. At the end of
the run the disc was experiencing very slow differential
precession, and extrapolation of the rate of twisting indicates
that this disc will achieve rigid-body precession with a
twist of $\sim 112^\circ$. This is the first indication in our
results that a disc which is predicted to be disrupted can
nonetheless generate internal stresses to enforce rigid precession
when the degree of twist becomes large. 

Models with $h=0.01$ and $R=10$ showed different behaviour,
largely as a result of the larger precession rate which is
induced for a larger disc, and suggest that the models run
with $R=8$ had outer radii slightly smaller than the natural 
tidal truncation radius.
The disc with $\alpha=h/2$, which supports bending waves,
was observed to undergo modest tidal truncation
and to break into two distinct
disc parts which precessed independently of one another,
an outcome which is similar to one obtained using an SPH
simulation by \cite{larwood} but for a disc with somewhat
different parameters.
A narrow rim at the edge of the disc, running between radii $r=$ 7 -- 10,
detached from the interior disc and precessed independently
at a faster rate. The inner disc part precessed at a 
rate similar to that observed in the $R=8$ case. Interestingly
we estimate that $\tau_W > \tau_P$ for this larger disc model,
such that the breaking of the disc is indeed expected. \\
The model with $\alpha=0.1$ propagates warps diffusively,
and was observed to differentially precess through
180 degrees, but maintained a very smooth twist profile
throughout the simulation. This disc is clearly disrupted through
strong differential precession, and the internal stresses are
insufficient for the disc to achieve rigid-body precession.
Over longer times this disc should become completely
twisted up by differential precession such that it loses all
semblance of a disc-like structure.

The results for the $R=10$ discs are interesting, as they indicate
two different modes by which a disc may break or be disrupted.
In the bending wave regime, our results suggest that disc disruption may occur
{\it via} a discontinuous change in the precession rate
at a particular radius, such that
the disc breaks into two distinct pieces which precess independently.
Within each separate disc-piece warp communication allows for rigid-body
precession. The detachment of the disc outer rim may in part
be influenced by the strong tidal interaction there, since 
non linear features and shocks may have the effect of reducing
the efficacy of warp communication.
In the diffusion regime, it appears that viscosity
is able to smooth out any discontinuities, and the disc is disrupted
through global differential precession of the disc annuli,
with a smooth twist profile being maintained across the disc radius. 

A significant result obtained for the $h=0.01$ models is
that these highly twisted discs align with the binary
orbit plane much more quickly than the thicker discs,
which align on the viscous timescale. Although we find that numerical
diffusion can cause an inclined disc to align with
the equatorial plane of the computational domain, this
occurs on much longer timescales than are observed.
We tentatively suggest that alignment of highly twisted
discs occurs on the precession time rather than the
viscous time, causing rapid alignment of very thin discs.

We have not considered different binary mass ratios, instead we restricted our study to the case of an equal mass binary.
As Eqs.~(\ref{free}) and (\ref{omega}) indicate, the free and mean precession rates scale linearly with the companion mass for a disc
of fixed outer radius.
Hence we would expect the differential precession timescale $\tau_P$ to be increased for less massive binary companions, and consequently 
the differential twist observed in our models should be reduced in such systems.

Our work has a number of astrophysical implications.
Most T Tauri stars are members of binary or
multiple systems, and in a number of these it is 
believed that the disc and orbit planes are misaligned.
One particular system for which there is a resolved
image of a disc which is misaligned from the binary plane is
HK Tau \citep{stapelfeldt}. In this system, the disc radius
is estimated to be 105 AU, and the projected binary separation
is approximately 3 times larger than this, suggesting that
the disc may be tidally truncated and undergoing strong
interaction with the inclined companion star. The images do not
show any signs that the disc is warped, which is
consistent with our $h=0.05$ models, whose aspect ratio
is probably appropriate for T Tauri discs.
Evidence for misaligned young binaries is also
provided by observations of precessing jets in star forming regions
\cite{eisloffel}. In one particular example,
\cite{Bally} suggest that the jet which appears to be 
launched by the young stellar object HH43* in the L1641 molecular cloud 
in Orion precesses with a period of about $10^4$ years. Applying models
to a disc with outer edge of 50 AU surrounding a solar type star, 
the rigidly precessing disc models presented in this work 
precess with a period of between 3.5 -- $5\times 10^4$ years, depending 
on the inclination angle considered. Thus the observed 
precession is consistent with that driven by a binary 
companion, with parameters close to those adopted in this work.
The long alignment timescales that we find for $h=0.05$ discs
suggests that a T Tauri disc will remain misaligned throughout
its lifetime, such that jets launched from the disc inner
regions will continue to precess for the duration of the
time when jet launching is active.

The thin $h=0.01$ disc models that we have computed are most likely
relevant to discs in X-ray binaries. The X-ray binary systems
SS433 \cite{margon} and Her X-1 \cite{boynton} are two examples
of systems thought to contain precessing discs,
with the evidence provided by the precessing jet of SS433
being particularly compelling. 
It is likely that the transfer of matter between the donor
star and compact object in these systems will cause disc
material to lie in the binary orbit plane initially, and a tilting
mechanism is required to move the disc out of this plane,
such as the Bardeen-Petterson effect \citep{bardeen}, a misaligned
dipole magnetic field associated with the central neutron star 
\citep{terqpap}, a disc wind \citep{schandl} or radiation pressure 
\citep{petterson-rad, pringle-rad}.
If our results on the rapid realignment of highly twisted discs are
correct, then this implies that a strong tilting mechanism 
which operates on short timescales will be required to maintain 
a misaligned disc which can be observed to precess.

Finally, we briefly consider the implications of our results
for the formation of planets in misaligned binary systems.
The early stages of planet formation are believed to
involve the growth of planetesimals {\it via} mutual collisions.
As the planetesimals grow in size, their interaction with
the disc through gas drag forces will decrease. Planetesimals
orbiting at different radii will have their own precession
frequencies, likely to be different to that of the disc.
Consequently, the growth of the planetesimals will eventually
cause them to develop orbits which are increasingly
inclined to the disc midplane. The effect of this is
likely to be an increase in relative motion between planetesimals,
which will affect the outcomes of mutual collisions,
and the rate at which planetary growth proceeds. 
A detailed examination of these issues will be presented
in a forthcoming paper.

\bibliographystyle{aa}
\listofobjects
\end{document}